\documentclass[prd,showpcs,amsmath,amssymb,nofootinbib,longbibliography,twocolumn,showpacs,notitlepage,superscriptaddress]{revtex4-1}

\usepackage[utf8]{inputenc}

\usepackage{latexsym}
\usepackage{epsfig}
\usepackage{graphicx,subfigure}
\usepackage[normalem]{ulem}
\usepackage{color}
\usepackage[usenames,dvipsnames,svgnames]{xcolor}
\usepackage{multirow}
\usepackage{hyperref}
\usepackage{numprint}
\usepackage{eucal}
\usepackage{amsmath}
\usepackage{amssymb}
\usepackage{comment}

\begin{document}

\begin{flushleft}
KCL-PH-TH/2024-25, IFT-UAM/CSIC-24-73 
 \end{flushleft}

\title{Investigating cosmic histories with a stiff era through Gravitational Waves
}

\author{Hannah Duval}
    \email{hannah.marie.d.duval@vub.be}
    \affiliation{Vrije Universiteit Brussel, Pleinlaan 2 - 1050 Brussel - Belgium}

\author{Sachiko Kuroyanagi}
    \email{sachiko.kuroyanagi@csic.es}
    \affiliation{Instituto de F\'isica Te\'orica UAM-CSIC, Universidad Aut\'onoma de Madrid, Cantoblanco 28049 Madrid - Spain}
    \affiliation{Department of Physics and Astrophysics, Nagoya University, 464-8602, Nagoya - Japan}

\author{Alberto Mariotti}
    \email{alberto.mariotti@vub.be}
    \affiliation{Vrije Universiteit Brussel, Pleinlaan 2 - 1050 Brussel - Belgium}

\author{Alba Romero-Rodríguez}
    \email{alba.romero-rodriguez@vub.be}
    \affiliation{Vrije Universiteit Brussel, Pleinlaan 2 - 1050 Brussel - Belgium}

\author{Mairi Sakellariadou}
    \email{mairi.sakellariadou@kcl.ac.uk}
    \affiliation{Theoretical Particle Physics and Cosmology Group, Physics Department, King's College, Strand, London WC2R 2LS - United Kingdom}

\date{\today}

\bibliographystyle{ieeetr}

\begin{abstract}
We investigate the potential of gravitational-wave background searches to constrain cosmic histories characterized by a stiff equation of state, preceded by a period of matter domination. Such a scenario leads to a characteristic peak in the primordial gravitational-wave spectrum originating from cosmological inflation. Assuming instant transitions between distinct epochs, which allows for an analytical treatment of the gravitational-wave spectrum, we perform a Bayesian inference analysis to derive constraints from the first three observing runs of the LIGO-Virgo-KAGRA Collaboration. Additionally, we consider a smooth transition, employing an axion-like particle physics model, and highlight the differences compared to the instant transition approximation. Finally, we forecast detection prospects for such a cosmic history through future gravitational-wave experiments.

\end{abstract}

\maketitle

\section{Introduction}
The LIGO-Virgo-KAGRA (LVK) Collaboration has already detected approximately one hundred transient gravitational-wave (GW) signals  arising from compact binary coalescences (CBCs) at the end of its third observing run
\cite{PhysRevX.11.021053, PhysRevX.13.041039}.
In addition, a significant amount of weak, unresolved sources are expected to contribute to a statistically random gravitational-wave background (GWB)~\cite{Allen_Romano_99, Romano_Cornish_2017}. 
The GWB can be of astrophysical (e.g., binary black holes (BBH), binary neutron stars (BNS), black hole-neutron star (BHNS) systems) or cosmological (e.g., first order phase transitions, topological defects, inflation ~\cite{Caprini:2018mtu}) origin.
On one hand, studies of the astrophysical GWB can provide useful information on aspects such as star formation rates, supernova explosions, the mass distribution of newly born black holes, and the mechanisms behind their growth~\cite{Regimbau:2011rp}.
On the other hand, studies of the cosmological GWB can provide information regarding early Universe mechanisms and theories beyond the Standard Model of particle physics whose energies are not accessible by accelerators~\cite{Caprini:2018mtu}.

In what follows, we focus on the primordial GWB spectrum sourced by inflation 
\cite{Starobinsky:1979ty,Rubakov:1982df}  
and explore how this GWB spectrum changes due to a non-conventional cosmological history with a period 
characterised by a stiff equation of state, motivated by particle physics theories beyond the Standard Model.
Traditionally, a model with a stiff epoch soon after inflation has been considered as an interesting example to enhance the inflationary GW amplitude at interferometer scale~\cite{Giovannini:1998bp,Peebles:1998qn,Giovannini:1999bh,Giovannini:1999qj,Giovannini:2008tm,Tashiro:2003qp,Figueroa:2018twl,Figueroa_2019}. While
in this paper, 
we consider another cosmological model in which a stiff epoch is preceded by a period of matter domination. 
This model is appealing 
because the traditional stiff domination model is severely constrained by the observational bounds by the cosmic microwave background (CMB) and Big Bang Nucleosynthesis (BBN), due to the increasing spectral amplitude towards high frequencies, which consequently restricts the parameter space accessible by GW experiments~\cite{Tashiro:2003qp,Figueroa:2018twl,Figueroa_2019}. In contrast, by including the matter domination epoch, the GW spectrum is suppressed at high frequencies, allowing for evasion of the CMB and BBN constraints. We investigate both instantaneous and smooth transitions from the matter epoch to the stiff epoch. The degree of smoothness in this transition is highly dependent on the specific model employed, and we illustrate this dependency using an axion-motivated model as an example~\cite{Co:2019wyp,Co:2021lkc,gouttenoire2022kination, Gouttenoire:2021wzu}.

The possibility of probing the cosmological history by properties imprinted in the primordial GWB spectrum, originating from scenarios with a non-standard cosmological history, has been investigated in several studies, see e.g.~\cite{Seto:2003kc,Smith:2005mm,Boyle_2008,Nakayama:2008ip,Nakayama:2008wy,Nakayama:2009ce,Kuroyanagi:2008ye,Kuroyanagi:2011fy,Kuroyanagi:2013ns,Kuroyanagi:2011fy,Kuroyanagi:2014qza,Kuroyanagi:2014nba,Bernal:2019lpc,Bernal:2020ywq,DEramo:2019tit, Li:2013nal, Li:2016mmc,Li:2021htg,Mishra:2021wkm, Battefeld:2004jh, DeAngelis:2024xtr, Haque:2021dha, Chakraborty:2023ocr}.
Observational constraints on such non-standard cosmological histories can be inferred by interpreting the results of the isotropic GWB search performed by the LVK Collaboration~\cite{LIGOScientific:2016jlg, LIGOScientific:2019vic,KAGRA:2021kbb}, from which constraints on the GWB with a power-law spectral tilt are derived. 
Although the spectrum of the GWB of this model is typically described by a power-law as long as we consider a narrow frequency range, this approximation can be violated when the frequency of observation corresponds to the transition phase from matter domination to stiff domination. 
In this paper, we present the first Bayesian inference search utilising LVK data from the first three observing runs (O1-O3) to constrain the parameters of an unconventional cosmological history, which includes a stiff epoch. 
Bayesian inference searches accommodate such exceptional cases where the power-law approximation cannot be utilised and also enable us to apply appropriate priors for the model parameters. Furthermore, we assess the detection prospects for future GW experiments, specifically for Advanced LIGO A+ \cite{Aplusdesign}, Laser Interferometer Space Antenna (LISA) \cite{colpi2024lisa}, Einstein Telescope (ET) \cite{Punturo:2010zz}, and a network of third-generation (3G) detectors, consisting of two ET-like detectors and two Cosmic Explorer (CE)-like detectors \cite{Hild_2011, borhanian2024listening}.

This article is organised as follows:
In Section \ref{sec:Basics} we review the basics of the primordial GWB sourced by inflation and how it is affected by a non-standard cosmological history.
In Section \ref{sec:Model building}, we build the GWB spectrum sourced by a period of stiff equation.
In Section \ref{sec:detection}, we study both current and future detection prospects by performing the Bayesian inference using the O1-O3 LVK data as well as mock data with future Advanced LIGO A+ sensitivity. Additionally, we present detectability estimates for next-generation detectors. We conclude with a short discussion in Section \ref{sec:discussion}.

\section{Primordial Gravitational waves in the framework of cosmology with a stiff equation of state}
\label{sec:Basics}

In standard cosmology and under the assumption of an inflationary phase, one considers that the period of inflation is followed by a radiation dominated (RD) era
and
a matter dominated (MD) era, which 
in turn transitions into
a dark energy dominated ($\Lambda$D) era. A common way to parameterise these different eras is through the  equation of state $w = P/\rho$,
where $P$ is the pressure and $\rho$ is the energy density of the fluid describing the Universe\footnote{Note that $w=1/3$ for RD, $w=0$ for MD, and $w = -1$ for inflation and for $\rm \Lambda D$.}.
While such a cosmological model is supported by observations of the
CMB and BBN,
there are still alternative scenarios possible between the end of inflation and the beginning of BBN.

In what follows, we study the primordial GWB spectrum induced by inflation and how it gets modified by an era with a non-standard equation of state (i.e., $1/3<w\leq1$) and how such a scenario can be identified and constrained through current and future GW experiments (see e.g.~\cite{Co:2021lkc,Li:2021htg,gouttenoire2022kination, Gouttenoire:2021wzu, Figueroa_2019} for previous studies).
We consider a model-independent parametrisation for the non-standard (referred to as \emph{stiff}) era, so that our analysis covers several particle physics models leading to such non-standard cosmologies.

\subsection{Primordial gravitational-wave spectrum from inflation}
In this section, we briefly discuss how the GW spectrum from inflation can be constructed, see e.g.~\cite{Guzzetti:2016mkm, Kuroyanagi:2008ye} for a more complete review. For this purpose, one considers GWs decomposed into Fourier modes and assumed to satisfy the transverse-traceless gauge conditions, propagating in a Friedmann-Lema\^itre-Robertson-Walker (FLRW) Universe and neglect anisotropic stresses that could affect their evolution. 
The GW equation of motion (EOM) reads 
\begin{equation}\label{eq:GWeqn}
h_{k}^{''\lambda}(\tau) + 2 \frac{a'(\tau)}{a(\tau)} h_{k}^{'\lambda}(\tau) + k^2 h_{k}^{\lambda}(\tau) =0~,
\end{equation}
where $k$ is the wave-vector of the Fourier mode, $\lambda$ is the polarisation index ($+$ or $\times$), 
$\tau$ is the conformal time defined as $d \tau= dt/a(t)$, and primes denote derivatives with respect to $\tau$.
Thus, the evolution of GWs in Eq.~\eqref{eq:GWeqn} is dictated only by the scale factor $a(\tau)$ and its time derivative.

During inflation, the GW modes oscillate quantum mechanically with
\begin{equation}
h_{k}^{\lambda}(\tau) \to \sqrt{16\pi G} \frac{e^{- i k \tau }}{a(\tau) \sqrt{2 k}} \,.
\end{equation}
Once they exit the 
Hubble radius $(aH)^{-1}$, with $H=(da/d\tau)/a^2$, 
%the Hubble parameter, 
and become super-Hubble 
their amplitude freezes.
At later times $\tau_{k}$, they re-enter the Hubble radius and become sub-Hubble,
so that $k = a(\tau_k) H(\tau_k)$. Then, their
amplitude decreases  with the scale factor as
\begin{equation}
h_{k}^{\lambda}(\tau) \propto \frac{1}{a(\tau)} e^{\pm i k \tau}~.
\end{equation}
The GW spectrum can be defined in terms of the GW energy density, $\rho_{\rm GW}$, and the critical energy density of the Universe today, $\rho_{c,0}$, as
\begin{equation}
\label{eq:gwdens}
    \Omega_{\rm GW}(f) = \frac{1}{\rho_{c,0}} \frac{d \rho_{\rm GW}}{d \ln (k)} \, ,
\end{equation}
where $\rho_{c,0} = 3H_0^2/(8 \pi G)$ and with $H_0$ representing the Hubble parameter today, defined as $H_0= 100 ~ h$ km s$^{-1}$ Mpc$^{-1}$. Throughout this paper we adopt a value for the dimensionless Hubble constant of $h=0.674$.
The GW energy density, expressed as a Fourier transform, reads
\begin{equation*}
    \rho_{\rm GW} = \frac{1}{32 \pi G} \int \frac{d^3 k}{(2\pi)^3} \frac{k^2}{a^2} \sum_{\lambda}|h_{k}^{\lambda}|^2~.
\end{equation*}
Therefore,
\begin{align}
\label{eq:gwdens2}
    \Omega_{\rm GW}(\tau, k) = \frac{k^2}{12 a^2(\tau) H^2(\tau)} \Delta_{\rm h}^2(\tau,k)\, ,
\end{align}
where the tensor power spectrum is defined as $\Delta_{\rm h}^2(\tau,k) \equiv \frac{k^3}{\pi^2} \sum_{\lambda}|h_{k}^{\lambda}|^2 $. 

For super-Hubble modes, the GW spectrum is nearly scale invariant, since the primordial tensor power spectrum is described by 
\begin{equation}
\label{eq:inflationspectrum}
 \Delta_{\rm h, inf}^2(\tau_i,k)
 = 
 \frac{2}{\pi^2} \Bigg(\frac{H_{\rm inf}}{m_{\rm p}} \Bigg)^2 \Bigg(\frac{k}{k_{\rm p}} \Bigg)^{n_{\rm t}}~.
\end{equation}
Note that $\tau_i$ is included to emphasise that we refer to times prior to Hubble radius re-entry. Here,
$n_t$ 
denotes the tilt of the spectrum, $k_{\rm p}$ is 
a pivot wavenumber, which is typically taken at the CMB scale, and $H_{\rm inf}$ is the Hubble energy scale
at which the $k_{\rm p}$ mode exits the Hubble radius~\cite{Ade_2021}. Within the paradigm of slow roll inflation, the consistency relation establishes a correlation between the spectral tilt and the tensor-to-scalar ratio: $n_t = -r/8$. 
The tensor-to-scalar ratio has an upper bound of $r < 0.036$ at $2 \sigma$, combining data from the CMB polarisation experiments Planck 2018, BICEP2, Keck Array and BICEP3~\cite{Ade_2021}.
This results in a $2 \sigma$ upper bound of $-n_t < 0.0045$.  In what follows,  we assume a scale invariant tensor power spectrum at super-Hubble scales, i.e. $n_t=0$ (we comment in the conclusion on the impact on our results of relaxing this assumption).  \\

For sub-Hubble modes, this formula is not valid and one needs to take into account a time-dependent factor, known as the transfer function
$T_{\rm h}^2(k,\tau)$~\cite{Turner:1993vb,Turner:1996ck}
\begin{equation}
\Delta_{\rm h}^2(\tau,k) = T_{\rm h}^2(\tau,k)\Delta_{\rm h,inf}^2(\tau_i,k)~.
\end{equation}
If the re-entry of the Hubble radius at $\tau_k$ occurs during a period in which the Universe is evolving with a 
constant equation of state, such that $a \propto \tau^{\alpha}$, 
with $\alpha$ defined as
\begin{equation}
\label{eq:alphadef}
\alpha \equiv \frac{2}{1+3 w}~,
\end{equation}
the transfer function reads
\begin{equation}
\label{eq:trasnfer1}
T_{\rm h}^2(\tau,k) = \frac{1}{2} \left( \frac{a(\tau_k)}{a(\tau)} \right)^2 \mathcal{A}_{\alpha} \,.
\end{equation}
Note that $\alpha$ should be evaluated at Hubble radius re-entry. The ratio of scale factors captures the redshift from the Hubble radius 
re-entry to later times $\tau$, while $\mathcal{A}_{\alpha}$ is a coefficient which accounts for a small difference in the GW evolution depending on the equation of state at Hubble radius 
re-entry, given by~\cite{Boyle_2008}
\begin{equation}
\label{eq:redshift}
     \mathcal{A}_{\rm \alpha} \equiv \frac{\Gamma^2(\alpha + 1/2)}{\pi} \left(\frac{2}{\alpha} \right)^{2 \alpha}~.
\end{equation}
We can thus infer the shape of the GW spectrum today which entered the Hubble radius 
in different cosmological eras, as
\begin{equation}
\label{eq:approxGW}
\Omega_{\rm GW} \sim k^{2(1-\alpha)} \qquad \text{with} \qquad 2(1-\alpha) = 2 \left(\frac{3w-1}{3w+1} \right)~.
\end{equation}
Hence,
for modes entering the Hubble radius  
during RD, the GW spectrum is flat,
while it increases or decreases with $k$, for $w>1/3$ or $w<1/3$, respectively.  Consequently, in the conventional cosmological history the inflationary GWB is flat.

However, in general, $T_{\rm h}^2(\tau,k)$ should be obtained by solving Eq.~\eqref{eq:GWeqn} for each mode, taking into account a possible non-trivial evolution of the scale factor. In particular, for modes which cross the Hubble radius while the equation of state of the Universe is changing, the evolution of the co-moving Hubble radius   cannot be described by a simple power law. 
As a result, the transfer function and the resulting GW spectrum should be determined by numerically solving Eq.~\eqref{eq:GWeqn} or analytically by matching the transition points assuming instant transitions. In the following, we employ both analytical and numerical approaches to address this scenario.

\subsection{Cosmological scenario with a stiff equation of state} \label{sec:Cosmological scenario with a stiff equation of state}

In this section, we consider a non-standard cosmological scenario, in which the Universe goes through an {\sl exotic} era characterised by a stiff equation of state $w_{\rm s}$, that precedes the standard RD era at BBN scales. Such a {\it stiff dominated} (SD) era would lead to a blue-tilted
inflationary GWB spectrum at frequencies higher than the one corresponding to BBN,
hence implying the potential for  detectable GW signatures
(see e.g~\cite{Giovannini:1998bp,Peebles:1998qn,Giovannini:1999bh,Giovannini:1999qj,Tashiro:2003qp,Giovannini:2008tm,Boyle_2008,Figueroa_2019} for
previous studies).

There are multiple models that motivate the presence of an SD period. For instance, in quintessence models, an SD era could occur just after inflation, characterised by an inflaton potential with a polynomial of degree $2 n$, leading to an equation of state with $w=(n-1)/(n+1)$~\cite{Turner:1983he}.
In such a scenario, the equation of state ranges from MD $w=0$ (for $n=1$) to the maximum allowed value for the equation of state during an SD era, which is referred to as kination, 
$w\simeq1$ (for $n \gg 1$). Alternatively, between the end of inflation and BBN, the energy density of the Universe can temporarily be dominated by a sector with an evolving equation of state, 
including an SD epoch. 
This phenomenon has been demonstrated
in axion-like models, as studied in e.g. ~\cite{Co:2021lkc,gouttenoire2022kination, Gouttenoire:2021wzu, Co:2019wyp}, where intermediate periods of MD and kination occur in between the radiation eras. 

\begin{figure}[h!]
\includegraphics[width=9cm]{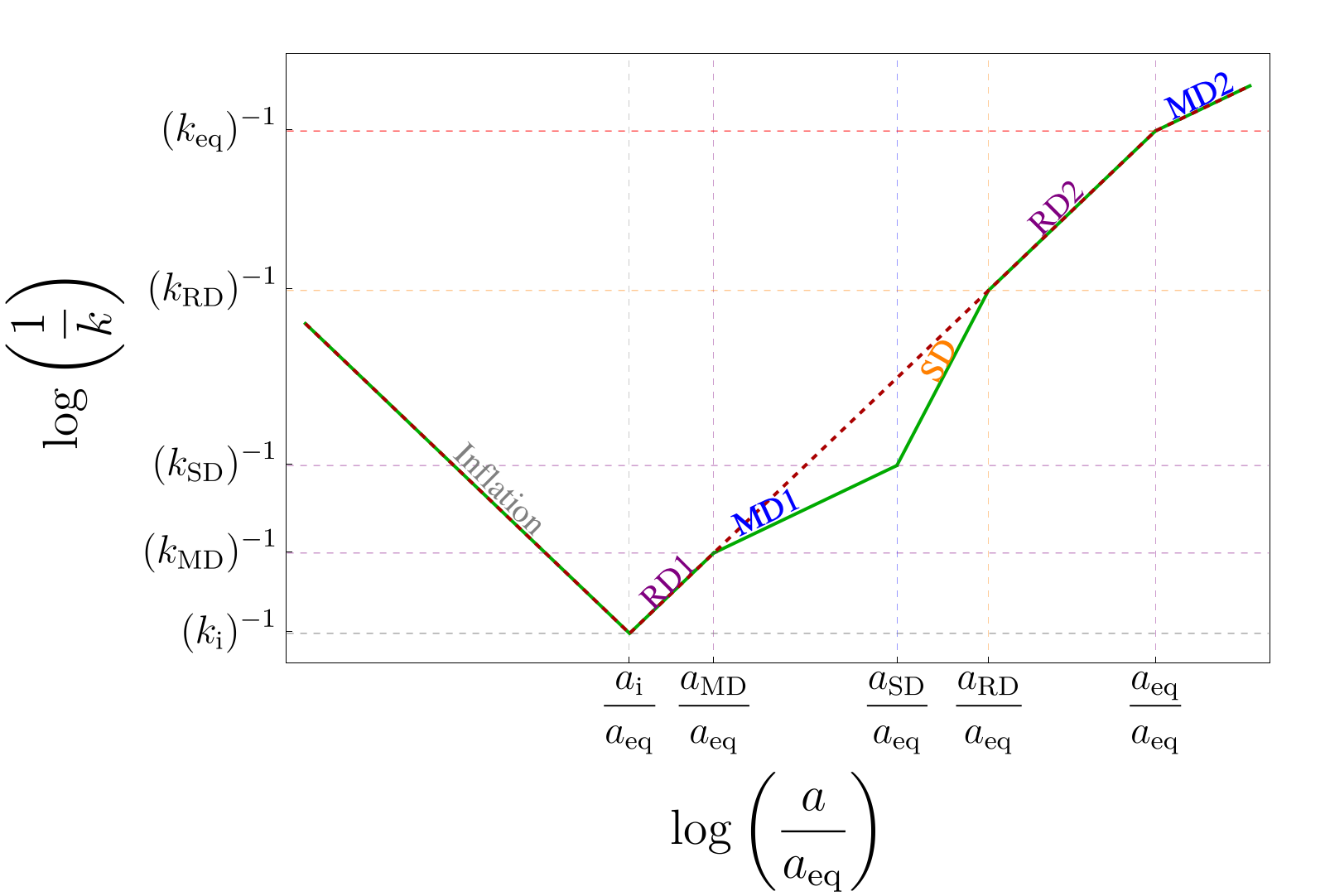}
    \caption{Schematic diagram illustrating the evolution of the co-moving Hubble radius $(aH)^{-1}$, represented in green, plotted against $a/a_{\rm eq}$. 
    Here, the equation of state during the stiff era is set to its maximum value: $w_{\rm s}=1$. 
    Additionally, the Hubble radius for the standard cosmological model is 
    plotted with
    a dashed dark red line.}
    \label{fig:Hubbleplot}
\end{figure}
In what follows, we work in a model-independent framework, where we consider a general Hubble evolution that can capture different scenarios. We consider a cosmological history in which the inflationary epoch precedes a first period of radiation domination (referred to as RD1), followed by a period of matter domination (referred to as MD1), which is then followed by a stiff dominated epoch (referred to as SD).  Subsequently, the SD era is followed by eras of radiation domination and matter domination (denoted as RD2 and MD2 respectively). 
The evolution of the co-moving Hubble radius as a function of the scale factor 
is 
shown in Fig.~\ref{fig:Hubbleplot},
and can be summarised as follows:
\small
\begin{align}
\label{eq:Hubblehorizon}
(a H)^{-1} \equiv (k)^{-1}\sim
    \begin{cases}
  a^{-1} & \mbox {for\ \  } a<a_{\rm i} \text{ (Inflation) } \\
a & \mbox {for\ \ } a_{\rm i}\leq a<a_{\rm MD} \text{ (RD1) } \\
 \sqrt{a} & \mbox {for\ \ } a_{\rm MD}\leq a<a_{\rm SD} \text{ (MD1) } \\
 a^{\frac{1}{2} (1 + 3 w_{\rm s})} & \mbox {for\ \ } a_{\rm SD}\leq
   a<a_{\rm RD} \text{ (SD) }\\
  a & \mbox {for\ \  }
   a_{\rm RD}\leq a<a_{\rm eq} \text{ (RD2) } \\
 \sqrt{a} & \mbox {for\ \ } a_{\rm eq}\leq a \text{ (MD2) }
\end{cases}
\end{align}
\normalsize
This scenario depends on several parameters: 
the Hubble scale of inflation $H_{\rm inf}$,
the value of the equation of state parameter during the stiff period $w_{\rm s}$, the wavenumber  of the 
RD1-to-MD1 transition $k_{\rm MD}$, the wavenumber of the MD1-to-SD transition $k_{\rm SD}$, and the wavenumber of the SD-to-RD2 transition $k_{\rm RD}$

We assume that during the MD1 era and the SD era that precede the {\sl standard} RD2 era, no entropy is injected. This can be seen in Fig.~\ref{fig:Hubbleplot}, from the fact that the red dashed line, which corresponds to the history in standard cosmology, connects the exotic RD1 era to the standard RD2 era. If we work in an entropy conservation scenario, the frequency at the RD1-to-MD1 transition can be related to the other transition frequencies and to the equation of state during the stiff era as
\begin{equation}
\label{eq:frdMax}
f_{\rm MD} = f_{\rm RD}^{\alpha_{\rm s}-1} f_{\rm SD}^{2-\alpha_{\rm s}}~,
\end{equation}
where $\alpha_{\rm s}$ denotes the index defined in Eq.~\eqref{eq:alphadef} during the stiff era.
The cosmological history is thus described by four independent parameters, $H_{\rm inf}, \alpha_{\rm s}, f_{\rm RD}$ and $f_{\rm SD}$, which will be used as basis in our parameter estimations performed in Section \ref{sec:detection}.

Assuming that the Hubble scale remains constant during inflation
and that there is no entropy injection between RD1 and RD2,
the frequency associated to the end of inflation
reads
\small
\begin{equation}
\label{eq:maxfmd}
f_{\text{i}} = 1.8 \times 10^8 \left(\frac{g_{s,{\rm RD}}}{106.75} \right)^{-\frac{1}{3}} \left(\frac{g_{*,{\rm RD}}}{106.75} \right)^{\frac{1}{4}} \left(\frac{H_{\rm inf}}{H_{\rm inf,max}} \right) \text{ Hz} \,,
\end{equation}
\normalsize
where $g_{s}$ represents the entropic degrees of freedom and $g_{*}$ the effective number of relativistic degrees of freedom
and RD denotes the value at the RD era well before the particles becomes non-relativistic,
here assumed to be equal to $g_{*,{\rm RD}}=g_{s,{\rm RD}}=106.75$. Additionally, we normalise the Hubble scale during inflation with $H_{\rm inf,max} = 5.12 \times 10^{13}$ GeV, which will be discussed in more detail in the next subsection. 

Our parametrisation actually encompasses several possibilities, where the most general one is the one given in Fig \ref{fig:Hubbleplot}. 
Assuming entropy conservation, $f_{\rm MD}$ is determined by the values of $f_{\rm SD}$, $f_{\rm RD}$ and $\alpha_{\rm s}$ as in Eq. \eqref{eq:frdMax}, and $f_{\rm MD, max}=f_i$ is the maximum allowed value within our scenario. When $f_{\rm MD}=f_{\rm i}$, the MD1 period starts right at the end of inflation.
From a phenomenological point of view we can consider values of $f_{\rm MD}$ larger than this, but in this case the cosmological history should be modified such that inflation directly connects to the MD era, possibly posing further model building challenges that go beyond our study.
We nevertheless include the case $f_{\rm MD} > f_{\rm i}$ in our parameter studies, since it can be easily re-interpreted in other scenarios.

\subsection{Asymptotic shape of the gravitational-wave spectrum}

In this section, we study the asymptotic behaviour of the inflationary GW spectrum in presence of the unconventional cosmological history. 

If the $k$-mode enters the Hubble radius in a regime far from transitions between two different eras, and the evolution of the Universe is governed by a constant equation of state, the scaling of the GWB amplitude can be calculated relatively easily.
Using Eqs.~\eqref{eq:gwdens2} - \eqref{eq:redshift}, in these regimes, the GWB spectrum scales as
\begin{widetext}
\begin{align}
\label{eq:Omegaasy}
    \Omega_{\rm GW}(f) =  \Omega_{\rm GW}|_{\rm plateau}^{(0)} \begin{cases}
        \mathcal{A}_{1}  & \mbox {for} \ \ \ \  f \ll f_{\rm RD} \\
        \mathcal{A}_{\rm \alpha_{\rm s}} \left(\frac{f}{f_{\rm RD}} \right)^{2(1-\alpha_{\rm s})} & \mbox {for} \ \ \ \  f_{\rm RD} \ll f \ll f_{\rm SD} \\
        \mathcal{A}_{2} \left(\frac{f_{\rm SD}}{f_{\rm RD}} \right)^{2(1-\alpha_{\rm s})}  \left(\frac{f_{\rm SD}}{f} \right)^2& \mbox {for} \ \ \ \  f_{\rm SD} \ll f \ll f_{\rm MD} \\
        \mathcal{A}_{1} \left(\frac{f_{\rm SD}}{f_{\rm RD}} \right)^{2(1-\alpha_{\rm s})}  \left(\frac{f_{\rm SD}}{f_{\rm MD}} \right)^2 & \mbox {for} \ \ \ \   f_{\rm MD} \ll f
    \end{cases}
\end{align}
\end{widetext}
where we have defined~\cite{Figueroa_2019}
\begin{equation}
    \Omega_{\rm GW}^{(0)}|_{\rm plateau} \equiv G_{\rm k} \frac{\Omega_{\rm rad}^{(0)}}{12 \pi^2} \left(\frac{H_{\rm inf}}{m_{\rm Pl}}\right)^2~,
\end{equation}
with $\Omega_{\rm rad}^{(0)} \approx 9 \times 10^{-5}$ and the reduced Planck mass $m_{\rm Pl} = 1/\sqrt{8 \pi G} \approx 2.44 \times 10^{18}$ GeV. Furthermore, we include a change in relativistic degrees of freedom through the factor $G_{\rm k}= [g_{\rm *,k}/g_{\rm *,0}] [g_{\rm s,0}/g_{\rm s,k}]^{\frac{4}{3}}$ which is frequency-dependent, 
where subscript $0$ denotes the values today ($g_{\rm *,0}=3.36$ and $g_{\rm s,0}=3.91$) and subscript $k$ represents the value when the corresponding mode enters the Hubble radius at $k=aH$.
Its full evolution can be found in~\cite{Watanabe:2006qe,Kuroyanagi:2008ye,Saikawa:2018rcs}.
The variation is small, ranging from $G_{\rm k} \sim 1$ for $f \sim 10^{-12}$ Hz to $G_{\rm k} \simeq 0.39$ for $f > 10^{-5}$ Hz.

The GBW spectrum has a peak around $f_{\rm SD}$. However, the exact shape of this peak cannot be accurately represented by the asymptotic behaviour described in Eq.~\eqref{eq:Omegaasy}. Instead, it should be determined by solving the GW EOM across the MD1-to-SD transition, either analytically or numerically. In particular, in Fig.~\ref{fig:Hubbleplot} the transitions between the different eras have been approximated as instantaneous. This approximation allows for a fully analytical derivation of the GW spectrum, achieved by matching the GW and its time derivative across different eras, following an approach previously adopted in~\cite{PhysRevD.37.2078, Figueroa_2019}. In more realistic models, the transition of the scale factor between two subsequent eras occurs smoothly, with the energy density of the Universe changing continuously as it transitions between components with different equations of state.
In the subsequent sections, we will consider both the instantaneous approximation and a smooth case to quantify their differences.

\subsection{Observational constraints}\label{sec:constraints}
Here, we describe observational constraints related to the GWB spectrum.
The inflationary scale can be 
written
in terms of the tensor-to-scalar ratio $r$, as $H_{\rm inf}= 2.7 \times 10^{14} ~ r^{1/2}$ GeV. 
The previously mentioned upper bound of $r < 0.036$ at $2 \sigma$, implies $H_{\rm inf} < H_{\rm inf,max} = 5.12 \times 10^{13}$ GeV, which is used throughout our analysis.
For the next generation CMB experiment LiteBIRD~\cite{Litebird_2022}, the upper bound on $r$ is expected to improve to  $r < 0.002$ at $2 \sigma$, leading to $H_{\rm inf} < 1.21 \times 10^{13}$ GeV. 

Furthermore, requiring that 
BBN should happen during an RD era and the SD epoch should end before,
we impose $f_{\rm RD} \geq f_{\rm BBN} \simeq 1.41 \times 10^{-11}$ Hz. Hence, $f_{\rm BBN} \leq f_{\rm RD} < f_{\rm SD} < f_{\rm MD} \leq f_{\rm i}$. In addition, there is a constraint on the GW energy density in terms of the effective number of neutrino species present in the thermal bath after $e^+ e^-$-annihilation $\Delta N_{\rm eff}$~\cite{Caprini_2018}
\begin{equation}
\label{eq:indirectbound1}
    \left(\frac{h^2 \rho_{\rm GW}}{\rho_{\rm c}} \right)\Big|_{\tau=\tau_0} \leq h^2 \Omega_{\gamma}^0 \frac{7}{8} \left(\frac{4}{11} \right)^{4/3} \Delta N_{\rm eff}~,
\end{equation}
where $h^2 \Omega_{\gamma}^0 = 2.47 \times 10^{-5}$ is the density parameter of photons today, and 
\begin{equation}
    \left(\frac{h^2 \rho_{\rm GW}}{\rho_{\rm c}} \right)\Big|_{\tau=\tau_0} = \int \frac{df}{f} h^2 \Omega_{\rm GW}(f)~.
\end{equation}
Considering a broken power law with a peak described in Eq.~\eqref{eq:Omegaasy}, we approximate the integral as~\cite{Caprini_2018}
\begin{equation}
    \left(\frac{h^2 \rho_{\rm GW}}{\rho_{\rm c}} \right)\Big|_{\tau=\tau_0} \approx \frac{1}{2 (1-\alpha_{\rm s})}h^2 \Omega_{\rm GW}(f_{\rm peak})~,
\end{equation}
where the right hand side is evaluated at the peak frequency $f_{\rm peak} \sim f_{\rm SD}$ (the precise peak location is found numerically in our analysis of Section \ref{sec:detection}).

The recent joint CMB+BBN analysis implies $\Delta N_{\rm eff} < 0.136 $ at $2 \sigma$~\cite{Yeh_2022}, that leads to
\begin{eqnarray}
h^2 \Omega_{\rm GW}(f_{\rm peak}) < (1-\alpha_{\rm s}) \cdot 1.53 \times 10^{-6}~.
    \label{eq:DeltaNeff}
\end{eqnarray}
The next generation of ground-based telescope experiments (CMB Stage-4) may improve the limit to $\Delta N_{\rm eff} < 0.02- 0.03$ at $1 \sigma$~\cite{abazajian2016cmbs4} 
Similar limits are expected from the Simons Observatory~\cite{SimonsObservatory:2018koc}.
By considering the $2 \sigma$-bound (taking $\Delta N_{\rm eff} < 0.04$) we obtain
\begin{eqnarray}
h^2 \Omega_{\rm GW}(f_{\rm peak}) <  (1-\alpha_{\rm s}) \cdot  4.49 \times 10^{-7}
    \label{eq:DeltaNefffuture}~.
\end{eqnarray}

Depending on the specific underlying model leading to the non-standard cosmological histories, further constraints could be set on the parameter space. 
For instance, for a particle physics model with a complex scalar field (radial mode plus axion), like the one we discuss in Appendix \ref{app:LogPot}, further constraints could arise from the mechanism responsible for damping the radial mode (see~\cite{gouttenoire2022kination, Gouttenoire:2021wzu, Co:2019wyp,Co:2021lkc}). In the spirit of a model-independent parametrisation of non-standard cosmological histories, we do not impose such constraints in our analysis and we restrict to the bounds which are directly set by the GW spectrum.

\section{Gravitational-wave spectrum}\label{sec:Model building}

In what follows, we derive the full inflationary GWB spectrum modified due to the cosmological history as described in Fig.~\ref{fig:Hubbleplot}.
As we mentioned before, the non-trivial parts of the GW spectrum correspond to the transitions between the different eras. 

In order to calculate
the full shape of the GWB spectrum including the transition regimes, we first assume that the transitions between the different eras occur instantaneously.
While this is
not a realistic assumption,
it has the advantage that one can solve the GW spectrum analytically. Next, we study a more physical scenario in which the MD1-to-SD transition is defined smoothly, motivated by axion models.

\subsection{Gravitational-wave spectrum in the instant transition scenario}
\label{sec:GWspectrum}
Here, we consider instantaneous transitions, characterising a sudden change between subsequent eras and occurring over a period significantly shorter than the Hubble scale at the transition moment. In this case, the equation of state can be described as a step-wise function, implying the following behaviour of the total energy density of the Universe throughout the different cosmological eras:
\begin{align}
\label{eq:rhototal}
    \rho_{\rm tot} = \begin{cases}
        \rho_i (\frac{a}{a_i})^{-4} &\text{ if \ \ } \tau_i \leq \tau \leq \tau_{\rm MD} \text{ (RD1) }\\
         \rho_{\rm MD} (\frac{a}{a_{\rm MD}})^{-3} &\text{ if \ \ } \tau_{\rm MD} \leq \tau \leq \tau_{\rm SD} \text{ (MD) }\\
        \rho_{\rm SD} (\frac{a}{a_{\rm SD}})^{-3(1+w_{\rm s})} &\text{ if \ \ } \tau_{\rm SD} \leq \tau \leq \tau_{\rm RD} \text{ (SD) }\\
         \rho_{\rm RD} (\frac{a}{a_{\rm RD}})^{-4} &\text{ if \ \ }  \tau \geq \tau_{\rm RD} \text{ (RD2) }
    \end{cases}
\end{align}
where ${a_i}= a(\tau_i)$, ${a_{\rm MD}}= a(\tau_{\rm MD})$ and so forth denote the scale factors at the corresponding transition times and
where the normalisation constants are equal to
\begin{align}
   \begin{cases}
        \rho_{\rm MD} &= \rho_* (\frac{a_{\rm MD}}{a_*})^{-4} \\
        \rho_{\rm SD} &= \rho_{\rm MD} (\frac{a_{\rm SD}}{a_{\rm MD}})^{-3}
        \\
        \rho_{\rm RD} &= \rho_{\rm SD} (\frac{a_{\rm RD}}{a_{\rm SD}})^{-3(1+w_{\rm s})}
    \end{cases}
    ~.
\end{align}
Given $\rho_{\rm tot}$, we then use the Friedmann-Lema\^itre equation to rewrite Eq.~\eqref{eq:GWeqn} as a Bessel equation. During a cosmological era, referred to as \textit{era 1}, which is followed by another \textit{era 2}, the solutions of the GW EOM are found to be
\begin{align}
\label{eq: GWsolint}
    h_{\rm 1}(y) &= A_{\rm  1} (\alpha_1 y)^{\frac{1}{2} (1-2\alpha_{\rm 1})} J_{\frac{1}{2} (-1+2\alpha_{\rm 1})}(\alpha_{\rm 1} y)\nonumber \\ 
    &+ B_{\rm 1} (\alpha_{\rm 1} y)^{\frac{1}{2} (1-2\alpha_{\rm 1})} Y_{\frac{1}{2} (-1+2\alpha_{\rm 1})}(\alpha_{\rm 1} y)~,
\end{align}
where we have omitted the polarisation index and where we introduced the variable $y\equiv k/(a H)$. Additionally, $J_{\nu}(x)$ and $Y_{\nu}(x)$ are Bessel functions of the first and second kind, respectively. 
The coefficients $A$ and $B$ can be found by matching the GW solutions of Eq.~\eqref{eq: GWsolint} and their time derivatives across different subsequent epochs.
In Appendix \ref{appendixmatching}, we give more details to this procedure and we work out the GW spectrum analytically.
This matching procedure ensures that the full GW spectrum contains information from all the considered eras. The full spectrum reads
\begin{eqnarray}
    \Omega_{\rm GW}(\tau_0,k) =&& \frac{4^{-2} G_k H_{\rm inf}^2 \Omega_{\rm rad}^{(0)} \Gamma \left(\frac{3}{2}\right)^2}{3 \pi ^3 \alpha_{\rm s} ^4 m_{\rm Pl}^2}\nonumber \\
    &&\times
   \left(\frac{f}{f_{\rm RD}}\right)^{-2 \alpha_{\rm s}} \left(\frac{f}{f_{\rm SD}}\right)^{2 (\alpha_{\rm s} -3)} \nonumber\\
   &&\times  {\cal F}(f, f_{\rm RD}, f_{\rm SD}, f_{\rm MD}, \alpha_{\rm s})~,
\end{eqnarray}
where ${\cal F}$ stands for a function which is a combination of Bessel functions. 

\begin{figure}[h!]
    \includegraphics[width=9cm]{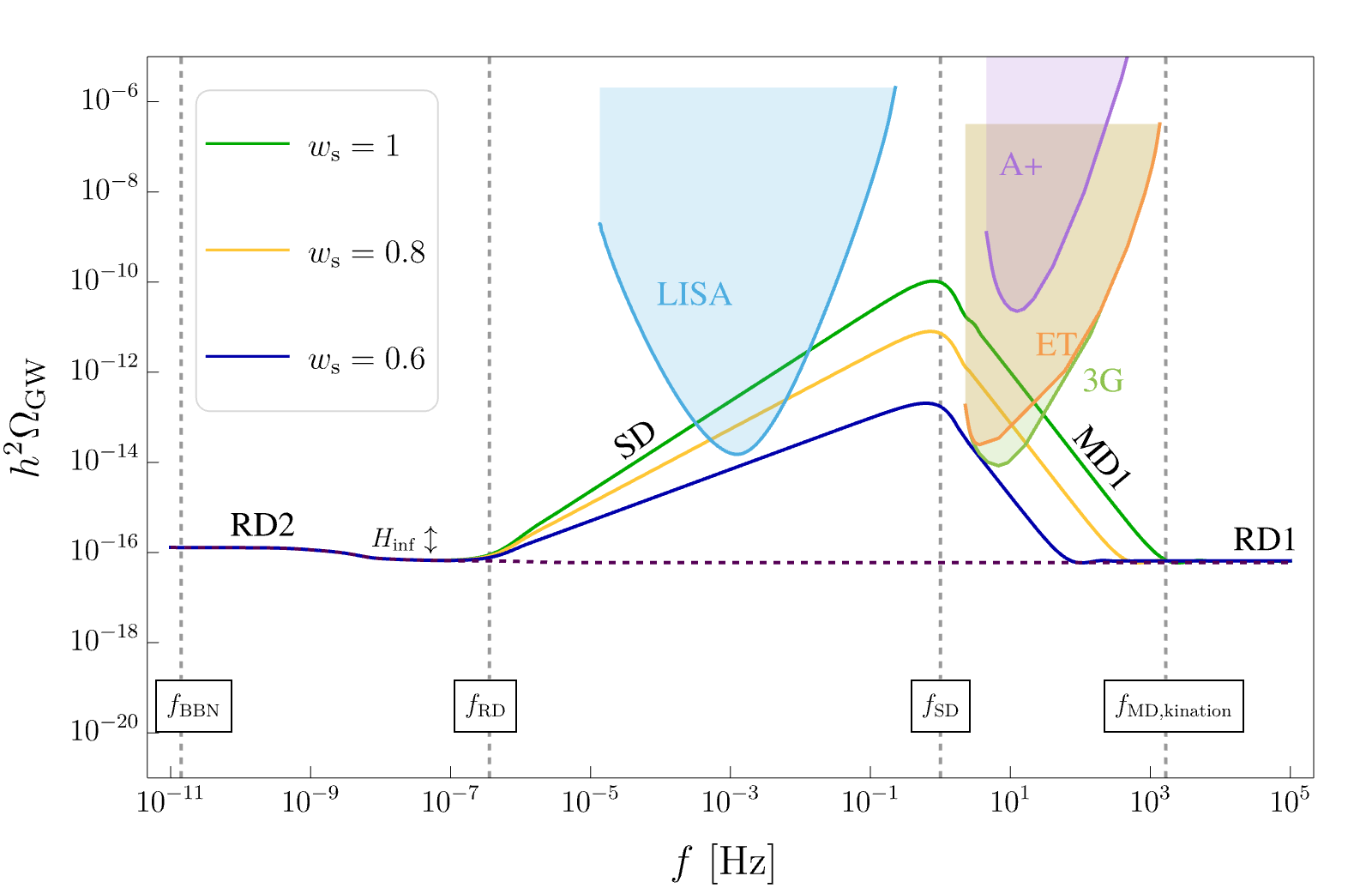}
 \caption{The analytically derived GWB spectrum 
 assuming instantaneous transitions for different values of $w_s$.
 The sensitivity curves for the future experiments that we will discuss later (see Table \ref{table:PI_curves}) are also shown. The parameters that influence the GW spectrum are chosen to be  
 $H_{\rm inf} = H_{\rm inf,max}$, $f_{\rm RD} = 3.6 \times 10^{-7}$ Hz, $f_{\rm SD} = 1$ Hz, and $f_{\rm MD}$ is fixed according to Eq.~\eqref{eq:frdMax}. The purple dashed line gives the standard inflationary GWB. The parameter $H_{\rm inf}$ influences the overall amplitude of the GWB spectrum, we denote this effect by the double arrow.}
    \label{fig:paramcomparisonplot}
\end{figure}

In Fig.~\ref{fig:paramcomparisonplot}, we show the analytic GWB spectrum for a few benchmarks and the sensitivity curves for Advanced LIGO A+, LISA, ET, and a network of 3G detectors (see Table \ref{table:PI_curves} and Section \ref{sec:futuredetectors} for the details of these future experiments and their sensitivity curves). The scaling of the analytical spectrum in the asymptotic regions is consistent with Eq.~\eqref{eq:Omegaasy}. 
As seen in the figure, $f_{\rm RD}$ sets the turning point between the flat inflationary spectrum and the increasing part and $f_{\rm SD}$ sets the GWB peak position, corresponding to the MD1-to-SD transition. We can also see that $\alpha_{\rm s}=2/(1+3w_{\rm s})$ determines the increasing slope of the GW spectrum, and that kination leads to the most promising scenario for detection.
The parameter $f_{\rm MD}$ is fixed
using
Eq.~\eqref{eq:frdMax}.
Finally, the inflationary scale $H_{\rm inf}$ sets the overall 
amplitude
of the GWB spectrum. Here and in what follows, we fix $H_{\rm inf}$ to its maximally allowed value: $H_{\rm inf}= H_{\rm inf,max}$, as discussed in Section \ref{sec:constraints}. 

Note that varying the parameters $H_{\rm inf}$ and $f_{\rm RD}$ while maintaining a constant value for the product $H_{\rm inf} f_{\rm RD}^{\alpha_{\rm s}-1}$ results in equivalent GW signals at frequencies above $f_{\rm RD}$.
Furthermore, as shown
in Fig.~\ref{fig:paramcomparisonplot},  GW signals at frequencies below or equal to 
$f_{\rm RD}$ are too weak to be detectable due to the upper bound on $H_{\rm inf}$. 
These two aspects lead to a degeneracy between $H_{\rm inf}$ and $f_{\rm RD}$. 

We thus note that our analysis with $H_{\rm inf}$ fixed to its maximum, can be re-interpreted in models with lower inflationary scales $H_{\rm inf}'$, by mapping $f_{\rm RD}$ to the frequency
\begin{eqnarray}
\label{eq:degeneracyhinffrd}
    f_{\rm RD}' &=& f_{\rm RD} \left( \frac{H_{\rm inf}'}{H_{\rm inf}}\right)^{\frac{1}{1-\alpha_{\rm s}}}\nonumber\\
    &=& f_{\rm RD} \left(\frac{H_{\rm inf}'}{5.12 \times 10^{13} \text{ GeV}}\right)^{\frac{1}{1-\alpha_{\rm s}}}~,
\end{eqnarray}
noting that $f_{\rm RD}'$ should be larger than $f_{\rm BBN}$.

\subsection{Gravitational-wave spectrum in the  smooth transition scenario}\label{sec:smoothinstant}

In this section, we study a smooth MD1-to-SD transition, for which we focus on the most promising scenario for a GWB detection, namely when the SD era is kination, with $w_{\rm s}=1$.  
As an example,
we consider an axion model previously discussed in~\cite{Co:2019wyp,Co:2021lkc,gouttenoire2022kination, Gouttenoire:2021wzu}, which we review in Appendix~\ref{app:LogPot}. This framework is based on a logarithmic potential for a complex scalar field, which splits in a radial mode and an axion. In such a model, the energy density at high temperature scales as matter, while at lower temperatures it scales as kination. The equation of state changes slowly, and the matter era is only reached asymptotically at high temperatures, implying that the peak in the GWB spectrum will be different 
from
the case of an instant transition. 

To facilitate a clear comparison, we first introduce the transfer functions to describe the GW spectrum for the case of kination with instantaneous transitions.
The full analytic expression derived 
in the previous subsection
can be fitted with the 
transfer functions as
\begin{equation}
\Omega_{\rm GW}(f) = \Omega_{\rm GW}^{(0)}|_{\rm plateau}  T_h^2(f,f_{\rm RD},f_{\rm SD}, f_{\rm MD})~,
\end{equation}
with
\begin{equation}
T_h^2 = T_{\rm RD}^2(f,f_{\rm RD}) T_{\rm SD}^2 (f,f_{\rm SD})T_{\rm MD}^2(f,f_{\rm MD})~,
\end{equation}
and where 
\begin{eqnarray}
\label{eq:TraInst}
    T_{\rm RD}^2(f,f_{\rm RD}) &=& 1-0.5 \left(\frac{f}{f_{\rm RD}} \right)^{2/3} + \frac{\mathcal{A}_{1/2}}{\mathcal{A}_{1}} \frac{f}{f_{\rm RD}}~,\nonumber\\
    T_{\rm SD}^2(f,f_{\rm SD}) &=& \left[ 1+0.55 \frac{f}{f_{\rm SD}}\right.
    \nonumber\\
   && \left.
    \ \  - 1.3 \left(\frac{f}{f_{\rm SD}}\right)^2 + \frac{\mathcal{A}_{1/2}}{\mathcal{A}_{2}} \left(\frac{f}{f_{\rm SD}}\right)^3 \right]^{-1} ~,\nonumber\\
    T_{\rm MD}^2(f,f_{\rm MD}) &=& 1-0.5 \frac{f}{f_{\rm MD}}  + \frac{\mathcal{A}_{1}}{\mathcal{A}_{2}} \left(\frac{f}{f_{\rm MD}}\right)^2~.
\end{eqnarray}
Eq. \eqref{eq:redshift} determines the redshift factors:
$\mathcal{A}_{1}=1$ in an RD era, $\mathcal{A}_{2}=9/16$ in a MD era and $\mathcal{A}_{1/2}=4/\pi$ during kination.
The asymptotics of Eq. \eqref{eq:TraInst} are consistent with the scaling in Eq.\eqref{eq:Omegaasy}.
The most interesting region for observations is the MD1-to-kination transition described by $T_{\rm SD}^2$, which characterises the GWB peak.
The GWB spectrum obtained by the transfer function fit shows relative deviations of at most $30\%$ (around the transition points) with respect to the full analytic expression of the instant transition case.

To describe the shape of the GWB corresponding to the smooth transition discussed in Appendix \ref{app:LogPot}, 
we first numerically solve for the GW spectrum
and then fit the resulting GWB peak
with a transfer function $T_{\text{log}}$ that should satisfy
\begin{eqnarray}
\label{eq:Tsmoothasy}
T_{\text{log}}^2 (f \ll f_{\rm SD})& \sim & 1~,\nonumber\\
\quad T_{\text{log}}^2(f \gg f_{\rm SD}) & \to &
\left(
\frac{f}{f_{\rm SD}}\right)^{-3} \frac{\mathcal{A}_2}{\mathcal{A}_{1/2}}~.
\end{eqnarray}
Eq.~(\ref{eq:Tsmoothasy}) provides a definition of $f_{\rm SD}$ for the smooth case, allowing a 
comparison between the smooth and instant transition cases (see Appendix \ref{app:LogPot} for details).
We find that the smooth MD1-to-kination transition can be well described  with the  fit
\begin{eqnarray}
    T_{\text{log}}^2(f,f_{\rm SD}) &&=\left[1+15 \left(\frac{f}{f_{\text{\rm SD}}} \right)^{5/2} \right.
    \\
   && \left.
    \ \ + 5 \left(\frac{f}{f_{\text{\rm SD}}} \right)^{2} +
 \frac{\mathcal{A}_{1/2}}{\mathcal{A}_{2}} \left(\frac{f}{f_{\text{\rm SD}}} \right)^{3}
 \right]^{-1}.\nonumber
\end{eqnarray}

\begin{figure}[h!]
    \includegraphics[width=9cm]{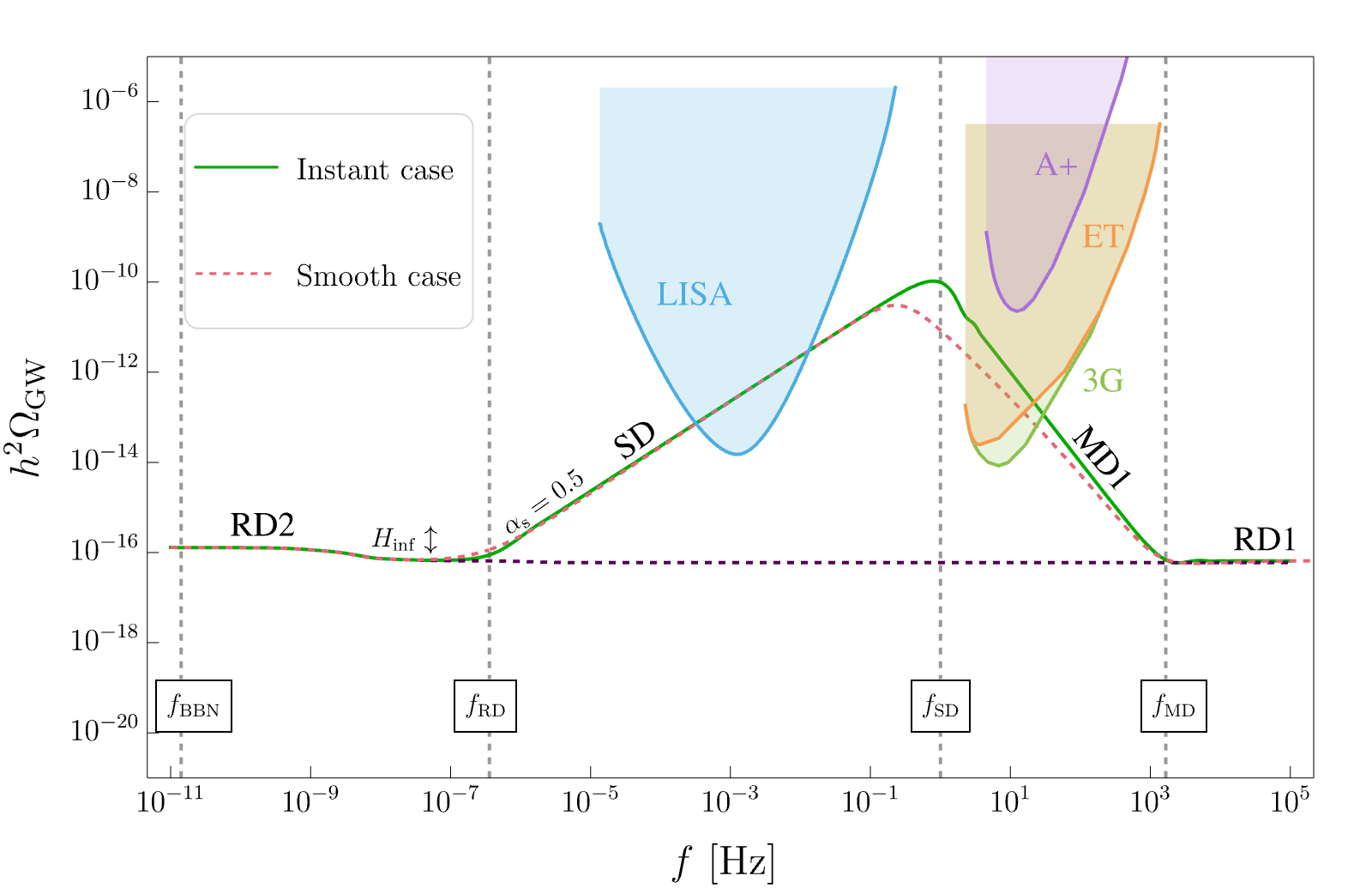}
 \caption{
 Comparison of the GW spectra derived for kination with $w_s=1$, assuming an instant MD1-to-kination transition (green) and a smooth one (dashed red). The parameters, $H_{\rm inf}$, $f_{\rm RD}$, $f_{\rm SD}$, $f_{\rm MD}$, are set in the same way as in Fig.~\ref{fig:paramcomparisonplot}. 
 The purple dashed line indicates the standard inflationary GWB spectrum.}
    \label{fig:smoothinstant}
\end{figure}

In Fig.~\ref{fig:smoothinstant} we compare the GWB spectrum for a smooth and an instant transition.
In both cases, the RD1-to-MD transition and the kination-to-RD2 transition 
are modelled as instantaneous ones, since their specific shape is beyond observational reach.
In terms of transfer functions, this corresponds to a GW spectrum 
\small
\begin{eqnarray}
\label{eq:smoothGWspectrum}
&&\Omega_{\rm GW}(f)\nonumber = \Omega_{\rm GW}^{(0)}|_{\rm plateau}  T_{\rm RD}^2(f, f_{\rm RD}) T^2_{\rm log}(f, f_{\rm SD}) T^2_{\rm MD} (f, f_{\rm MD})~.\nonumber
\end{eqnarray}
\normalsize
The only transition that differs is the matter-kination one, around $f_{\rm SD}$.

We subsequently investigate the 
impact of modelling the transition as instantaneous or
smooth regarding detection prospects, and conclude that the type of transition has minor impact on the parameter region reach of
GW experiments. 
Consequently, we infer that the results obtained in the instantaneous transition case remain applicable for the smooth transition case. Hence, in the next section we will solely conduct a data analysis on the instantaneous transition case.

\section{Detection prospects}\label{sec:detection}
In this section, we present the detection prospects for current and future detectors. Although current ground-based detectors have not yet made a GWB detection, we can place stringent constraints on the parameter space of our model using O1-O3 LVK data. 
In what follows, we first discuss the general methodology and then perform a Bayesian inference search considering both a CBC background and 
an inflationary GWB signal enhanced by a stiff epoch
{with the current LVK data.
We then investigate the detection prospects of future experiments 
assuming Advanced LIGO A+ sensitivity.

\subsection{Bayesian inference formalism used in Gaussian, stationary, unpolarised and isotropic gravitational-wave background  searches}\label{sec:PE}
We use O1-O3 LVK data to constrain the exotic cosmological model with a stiff epoch ~\cite{LIGOScientific:2016jlg,LIGOScientific:2019vic,KAGRA:2021kbb,LIGOScientific:2019lzm,KAGRA:2023pio}. We perform a Bayesian inference search as described in~\cite{PhysRevLett.109.171102, PhysRevX.7.041058, PhysRevD.102.102005, Abbott_2021}. The Gaussian likelihood for the search of an isotropic unpolarised GWB reads
\begin{equation} \label{eq:PE_likelihood}
p(\hat{C}^{IJ}(f)|\theta) \propto \exp \!\!\left[ -\frac{1}{2} \sum_{IJ} \sum_f \left( \frac{\hat{C}^{IJ}(f) - \Omega_{\rm M}(f|\theta)}{\sigma_{IJ}(f)}\right)^2  \!\right]\!,
\end{equation}
where quantity $\hat{C}^{IJ}(f)$ is the cross-correlation estimator of the GWB, computed using data from detectors $I$ and $J$, and $\sigma_{IJ}(f)$ its associated variance~\cite{PhysRevD.59.102001, Renzini_2023, Romano_Cornish_2017}.
The sums run over the detector baselines  $IJ$ and over the frequencies $f$.  The parameter $\Omega_{\rm M}(f|\theta)$ describes the GWB model, where $\theta$ represents the model parameters. 
In our case, $\Omega_{\rm M}(f|\theta) = \Omega_{\rm SD}(f) + \Omega_{\rm cbc}(f) $, where $\Omega_{\rm SD}(f)$ represents the inflationary GWB enhanced by a stiff epoch and $\Omega_{\rm cbc}(f)$ accounts for the astrophysical GWB from unresolved CBCs. Within the frequency range of interest, the latter takes the form~\cite{PhysRevX.7.041058}
\begin{equation}
    \Omega_{\rm cbc}(f)=\Omega_{\rm ref}\left(\frac{f}{f_{\rm ref}}\right)^{2/3}~,
\end{equation}
where $f_{\rm ref}$ is a reference frequency, set to $f_{\rm ref}=25$ Hz and $\Omega_{\rm ref}$ is the amplitude of the CBC background at the reference frequency. \\
\noindent Furthermore, we will quote a Bayesian upper limit (UL) at $95 \%$ confidence level (CL). For a parameter $a$ and data $d$, this is defined as~\cite{Romano_Cornish_2017}
\begin{equation}
    \rm Prob(0<a<a^{95 \% UL}|d) = 0.95.
\end{equation}
In order to compare different models, we make use of Bayes factors. Given two models $\mathcal{M}_a$ and $\mathcal{M}_b$, the Bayes factor can be computed as
\begin{equation}
    \mathcal{B}^a_b (d) = \frac{p(d|\mathcal{M}_a)}{p(d|\mathcal{M}_b)},
\end{equation}
where $p(d|\mathcal{M}_i)$ denotes the evidence for model $\mathcal{M}_i$ ($i=a,b$). Additionally, one can compute $\ln(\mathcal{B}^a_b (d))$:
\begin{equation}
    \ln(\mathcal{B}^a_b (d)) \equiv \ln(\mathcal{B}^a_{\rm Noise} (d)) - \ln(\mathcal{B}^b_{\rm Noise} (d)) ~.
\end{equation}
Positive values of $\ln(\mathcal{B}^a_b (d))$ imply preference for
model $\mathcal{M}_a$ over $\mathcal{M}_b$.
All the results presented in the following are produced using the Python packages {\tt pygwb}~\cite{Renzini_2023} and {\tt Bilby} ~\cite{Ashton_2019} using the {\tt Dynesty} sampler. 

\subsection{Constraints on the stiff epoch using LVK gravitational-wave background search data}\label{sec: PE_O3}
Here, we employ the formalism described in Section~\ref{sec:PE} to constrain the model parameters characterising the full unconventional cosmological history with a stiff epoch.
Using the GW data obtained from the initial three observing runs of the LVK Collaboration, we do an analysis considering both a CBC background and 
an inflationary GWB enhanced by a stiff epoch, from now on referred to as SD +CBC model. Furthermore, we perform Bayesian inference searches using mock data consisting of the expected noise of the Advanced LIGO A+ sensitivity and injecting an SD + CBC GWB.

As seen previously, the parameter space is given by $\theta = (\Omega_{\rm ref}, H_{\rm inf}, f_{\rm MD},  f_{\rm SD}, f_{\rm RD}, \alpha_{\rm s}
)$. 
In the parameter estimation, we fix the parameters $H_{ \rm inf}$ and $f_{\rm MD}$ by using delta function priors centred around their respective maximum values.
In particular, in order to encompass generic scenarios (see discussion in Section \ref{sec:constraints}), we do not impose the entropy conservation condition in Eq. \eqref{eq:frdMax} and we fix $f_{\rm MD}$ to $f_{\rm i}$. With this configuration, the GW spectrum in the LVK sensitivity band is determined by $f_{\rm RD}$ and $f_{\rm SD}$ and, at frequencies higher than $f_{\rm SD}$, the spectrum continues to decrease with $\Omega_{\rm GW} \propto f^{-2}$ until $f_{\rm i}$. Since this high-frequency part is not detectable with the LVK sensitivity, the choice of $f_{\rm MD}$ does not affect our analysis.
The priors chosen for the remaining parameters are given in Table~\ref{table:priors}.
\begin{table}[h!]
\centering
\begin{tabular}{||c| c ||} 
 \hline
 Parameters $\theta$ & Priors \\ 
 \hline\hline
 $\Omega_{\rm ref}$  & \rm LogU($10^{-13}, 10^{-5}$) \\ 
 \hline
 $f_{\rm RD}$ [Hz] & \rm LogU($10^{-10}, 10^{-5}$) \\
 \hline
 $f_{\rm SD}$ [Hz] & \rm LogU($10^{-3}, 10^{6}$) \\
 \hline
 $\alpha_{\rm s}$ & \rm U($0.5, 1$) \\
 \hline
\end{tabular}
\caption{Priors applied to the Bayesian inference. They are categorised as either uniform ($\rm U$) or uniform in logarithmic scale ($\rm LogU$). The prior set for $\Omega_{\rm ref}$ is derived from previous estimates of the CBC background contribution~\cite{Abbott_2018}. The priors for $\alpha_s$ are theoretically determined, based on the minimum and maximum values of $w_s$: $1/3 \leq w_{\rm s} \leq 1$. Considering that the RD2 epoch precedes the SD epoch and that BBN should happen during RD2, it follows that $f_{\rm BBN} \leq f_{\rm RD} < f_{\rm SD}$. 
The prior ranges for $f_{\rm RD}$ and $f_{\rm SD}$ are chosen wide enough to make sure that a subset of them leads to signals detectable within the LVK frequency band. We checked that extending the prior ranges do not significantly modify the resulting posterior distributions.
}
\label{table:priors}
\end{table}
\subsubsection{O1-O3 LVK data}
In Fig.~\ref{fig:PE_O3_1}, we show the results from a Bayesian inference search, as described in Section \ref{sec: PE_O3}, illustrating the resulting posterior distributions. The top panel shows the results for a general SD era, where $\alpha_{\rm s}$ is varied, while the lower panel discusses a kination scenario, for which a delta prior centred around $\alpha_{\rm s}=0.5$ is set. 
\begin{figure}[h!]
    \includegraphics[scale=0.35]{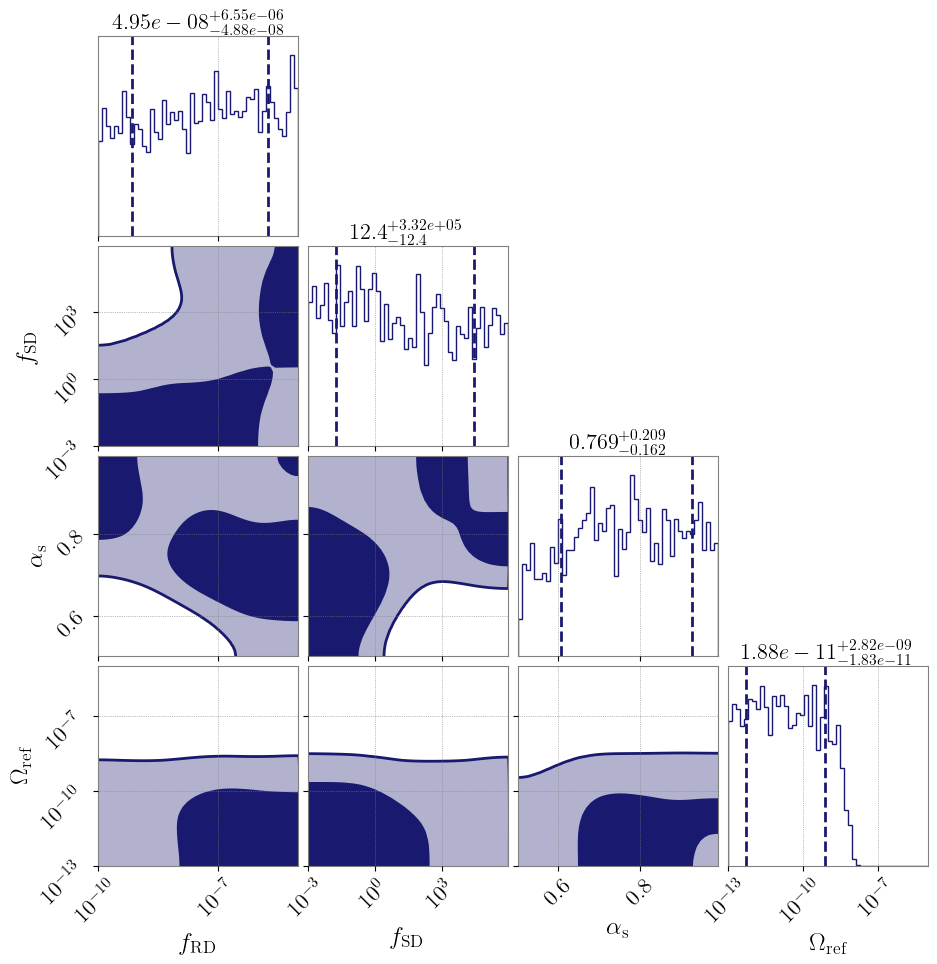}
    \includegraphics[scale=0.35]{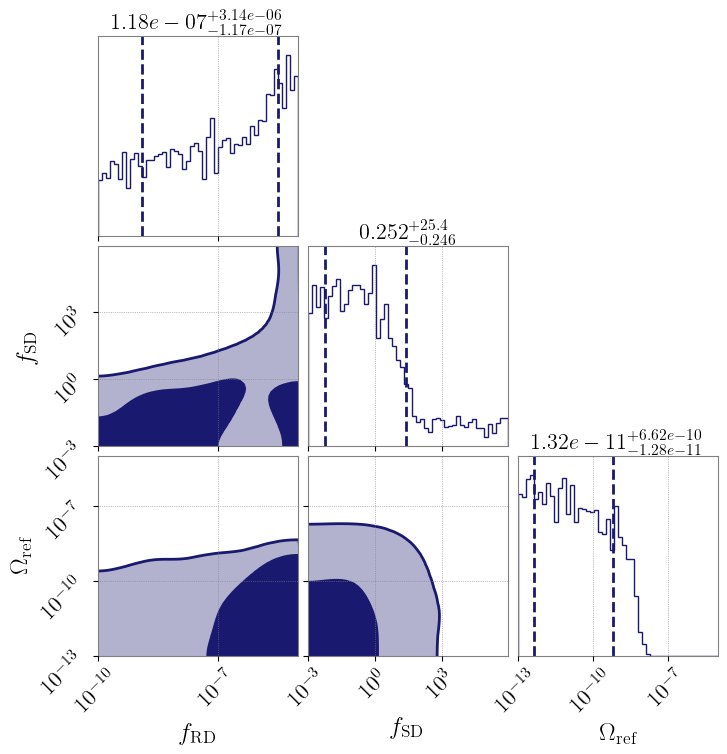}
    \hspace{-1cm}
    \caption{Posterior distributions from Bayesian inference assuming an SD+CBC model. 
    Contour regions in dark blue correspond to $1 \sigma$ CL and those in lighter blue to $2 \sigma$ CL. 
    In the upper part of the triangular plot, the 1-dimensional marginalised posterior distribution is shown and the dashed vertical lines represent the $1 \sigma$ CL. Furthermore, the median and errors of the posteriors are shown on the top.
    In the top panel, all parameters vary, while in the bottom panel we show the full kination case where $\alpha_{\rm s}$ is fixed to $\alpha_{\rm s}=0.5$. }
    \label{fig:PE_O3_1}
\end{figure}
\noindent The posterior distributions depicted in Fig.~\ref{fig:PE_O3_1} allow us to establish $95\%$ CL ULs on $\Omega_{\rm ref}$. Specifically, these limits are determined to be around $2.8 \times 10^{-9}$ for both cases of varying and fixing $\alpha_{\rm s}$. These findings align with previous ULs on the amplitude of the CBC background, which were reported to be $3.4 \times 10^{-9}$ at $25$ Hz~\cite{Abbott_2021}. Moreover, the contours representing the $1 \sigma$ CL for $f_{\rm RD}$, $f_{\rm SD}$, and $\alpha_{\rm s}$ are shaded in dark blue, while those corresponding to the $2 \sigma$ CL are presented in lighter blue. The vertical dashed lines in the histograms of the posterior distributions denote the $1 \sigma$ CL.
In the full kination case (as shown in the lower panel of Fig.~\ref{fig:PE_O3_1}), a $95\%$ CL UL can be established for $f_{\rm SD}$, yielding $21716$ Hz and a $95\%$ CL lower limit\footnote{While typically ULs are determined using O1-O3 data, in this instance, we discuss a lower limit. This is because lower values of $f_{\rm RD}$ correspond to higher peaks in the GW signal.} 
for $f_{\rm RD}$ of $2.35 \times 10^{-10}$ Hz. This suggests a preference for lower values of $f_{\rm SD}$ and higher values of $f_{\rm RD}$.
The first three observing runs of the LVK Collaboration enable us to constrain certain regions in parameter space. Specifically, within the ($f_{\rm RD}, f_{\rm SD}$) space, small values of $f_{\rm RD}$ and large values of $f_{\rm SD}$ are excluded at $2 \sigma$ CL.
This exclusion region corresponds to signals that would have been detected with O1-O3 data.

In the general case (top panel of Fig.~\ref{fig:PE_O3_1}), a portion of the $(f_{\rm RD}, \alpha_{\rm s})$ parameter space is excluded at the $2 \sigma$ CL, corresponding to small values of $\alpha_{\rm s}$ and $f_{\rm RD}$. Similarly, another portion of the $(f_{\rm SD}, \alpha_{\rm s})$ parameter space is also excluded at the $2 \sigma$  CL for small values of $\alpha_{\rm s}$ and large values of $f_{\rm SD}$. Moreover, within the $(f_{\rm RD}, f_{\rm SD})$ space, $2 \sigma$ CL exclusions apply to small values of $f_{\rm RD}$ and large values of $f_{\rm SD}$. Notably, this exclusion region is comparatively smaller than that observed for the full kination case, as variability in $\alpha_{\rm s}$ in the top panel results in weaker signals.
The posterior distribution of $\alpha_{\rm s}$ exhibits a preference for larger values. Additionally, the posterior distribution of $f_{\rm RD}$ exhibits a slight increasing trend, while the one of $f_{\rm SD}$ shows a slight declining trend, in agreement with our expectations from the full kination case.

We also compute the Bayes factors comparing the hypothesis of data containing an SD+CBC signal versus noise only. In the case of kination, the Bayes factor is $\ln(\mathcal{B}_{\rm Noise}^{\rm kination+CBC}) = -1.11$. In the case of a varying $\alpha_{\rm s}$, $\ln(\mathcal{B}^{\rm SD+CBC}_{\rm Noise}) = -0.65$, indicating no evidence for an SD+CBC signal in the data. Similarly, we obtain Bayes factors between the hypothesis of data containing a CBC background versus noise only: $\ln(\mathcal{B}^{\rm CBC}_{\rm Noise}) = -0.62$, showing no evidence for a CBC background in the data. Moreover, we compute the Bayes factors comparing the hypothesis of data containing a CBC background versus an SD+CBC signal. For the kination case, we obtain $\ln(\mathcal{B}_{\rm CBC}^{\rm kination+CBC}) = -0.49$, indicating a preference for a CBC background in the data. In the case of varying $\alpha_{\rm s}$, $\ln(\mathcal{B}_{\rm CBC}^{\rm SD+CBC}) = -0.03$, showing a slight preference for a CBC background only.

In summary,
we do not find evidence in the data for a CBC background nor for a signal coming from a stiff epoch. We are therefore able to set $95\%$ CL ULs on some of the parameters describing the exotic cosmological history.

\subsubsection{Recovery of injections using Advanced LIGO A+ sensitivity}
\label{sec:O5}
In this section, we study some possible challenges that may arise with the detection of an SD+CBC GWB. 
To address these challenges, 
we again perform a Bayesian inference search by using mock data consisting of the expected noise of the Advanced LIGO A+ sensitivity, where we further inject two GWB signals: one originating from a CBC background and the other from the exotic cosmological history with instant transitions. 
The CBC background amplitude is set to $\Omega_{\rm ref} = 7 \times 10^{-10}$, which corresponds to the central value 
of the estimated astrophysical GWB as derived 
in the O1-O3 LVK analysis
\cite{Abbott_2021}. 

A first inherent challenge arises, which has been  discussed in \cite{Martinovic_2021},
when employing Bayesian inference searches to simultaneously reconstruct the GWB from CBCs and a GWB originating from cosmological models. These challenges persist within our framework as well.
In particular, our model predicts a broken power-law spectrum, where the location of the peak corresponds to $f_{\rm SD}$. The results of the parameter estimation significantly depend on the location of this peak for the injected signal.
If $f_{\rm SD}$ is relatively large ($f_{\rm SD} \gtrsim 50$ Hz), the resulting GWB signal resembles, within the LVK frequency range,
a power-law spectrum with positive spectral index, mimicking the spectrum of the CBC background. Consequently, the Bayesian analysis cannot accurately recover the CBC parameter $\Omega_{\rm ref}$ and the parameters of the cosmological model simultaneously. 

Conversely, if the value of $f_{\rm SD}$ falls below the frequency range of LVK, the resulting GWB signal appears as a power law with negative spectral index within this range. This enables to better resolve the GWB signal from the stiff epoch and the one from the CBC background. 
We perform a Bayesian inference on this second case.
The cosmological parameters are 
listed
in Table~\ref{table:benchmarks} and
we employ the same priors as before, given in Table \ref{table:priors}.
 
\begin{table}[h!]
\centering
\begin{tabular}{||c| c ||} 
 \hline
 Parameters $\theta$ & Benchmark values \\ 
 \hline\hline
 $f_{\rm RD}$ [Hz] & $2.5 \times 10^{-10}$ \\
 \hline
 $f_{\rm SD}$ [Hz] & $5$ \\
 \hline
 $f_{\rm MD}$ [Hz] & $1.8 \times 10^8$ \\
 \hline
 $H_{\rm inf}$ [GeV] & $5.12 \times 10^{13}$ \\
 \hline
 $\alpha_{\rm s}$ & $0.588$ \\
 \hline
\end{tabular}
\caption{Benchmark values used for the injected signal in the Bayesian inference search. These values were selected to ensure detectability of the signal by the expected Advanced LIGO A+ sensitivity, and to have a stiff signal behaving differently than the CBC signal in LVK frequency range. For $w_{\rm s}$, we choose a value of  $0.8$, corresponding to $\alpha_{\rm s} = 0.588$. 
For $f_{\rm MD}$ and $H_{\rm inf}$, we adopt their maximum values allowed by observational constraints. A relatively small value is assigned to $f_{\rm RD}$ based on the fact that smaller values give rise to stronger GWB signals.
}
\label{table:benchmarks}
\end{table}
The resulting posterior distributions are illustrated in Fig.~\ref{fig:PE_injection_all}.
The parameter $\Omega_{\rm ref}$ is not successfully recovered due to the weaker strength of the CBC signal compared to the stiff signal ($\Omega_{\rm SD}(25 \text{ Hz })= 1.9 \times 10^{-9}$ and $\Omega_{\rm cbc}(25 \text{ Hz })= 7 \times 10^{-10}$), but we nevertheless obtain an upper bound on the CBC background amplitude.
The stiff parameter $f_{\rm SD}$ is recovered within $1 \sigma$.
However, the recovery of the other two stiff parameters, $f_{\rm RD}$ and $\alpha_{\rm s}$, is less successful due to their \textit{degeneracy} in determining the overall GWB amplitude in the LVK frequency range, an issue which explains the elongated 2D contour regions. 
This poses a second challenge that arises in our model, which has multiple signal degeneracies.
While the degeneracy between $H_{\rm inf}$ and $f_{\rm RD}$ has been previously addressed in Section \ref{sec:Cosmological scenario with a stiff equation of state}, we now further explore additional ones\footnote{While we employ the term \textit{degeneracies}, it is important to note that the ones we discuss in this subsection might appear as such, but they are not true degeneracies. This resemblance arises due to the inability of LVK to distinguish accurately between slopes.} within the LVK frequency band.
\begin{figure}[h!]
    \includegraphics[width=9cm]{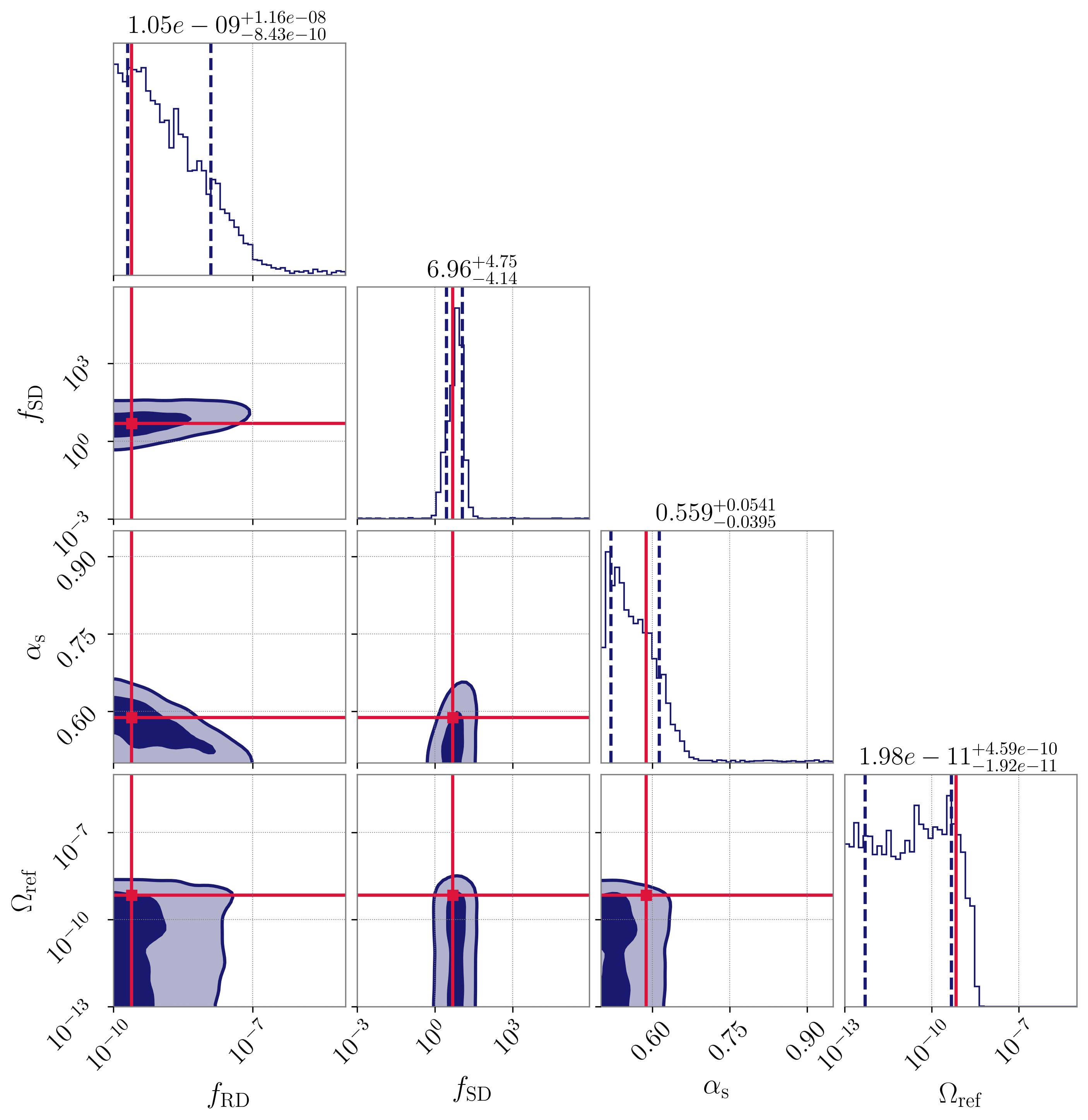}
    \caption{Posterior distributions from running a Bayesian analysis assuming an SD+CBC model, using mock data at Advanced LIGO A+ sensitivity. 
    The injected values are indicated by the red lines.    
      }
    \label{fig:PE_injection_all}
\end{figure}

To further clarify this, we 
focus on the degeneracy 
between $f_{\rm RD}$ and $\alpha_{\rm s}$. 
Again, an injection is made with parameter values given in Table ~\ref{table:benchmarks}, but we now only set varying priors on $f_{\rm RD}$ and $\alpha_{\rm s}$. Furthermore, we do not include the CBC background.
The results are shown in Fig. \ref{fig:PE_degeneracy_each},
where the degeneracy in the 
$(f_{\rm RD},\alpha_{\rm s})$
plane is clearly visible in the blue contour regions.
In the same figure, we include the theoretically expected degeneracy depicted in yellow. 
For this purpose, 
we examine the injected GW amplitude at the frequency of highest sensitivity ($f \approx 35$ Hz),
which is equal to $\Omega_{\rm SD}(35 \text{ Hz }) \simeq 9.6 \times 10^{-10}$.
We then employ the analytic formula for the GWB to identify the combinations of
$f_{\rm RD}$ and $\alpha_{\rm s}$ 
that produce a GWB spectrum with $\Omega_{\rm SD}(35 \text{ Hz }) \simeq 9.6 \times 10^{-10}$ .
This determines a degeneracy line that is shown in yellow in 
Fig. \ref{fig:PE_degeneracy_each}.
From this figure, one can see that the analytically derived degeneracy corresponds to the retrieved 2D contour regions. We have verified that the other parameter degeneracies of the cosmological model seen 
in Fig. \ref{fig:PE_injection_all} can be reconstructed using similar analytic arguments.

\begin{figure}
    \hspace{-1cm}
    \includegraphics[scale=0.45]{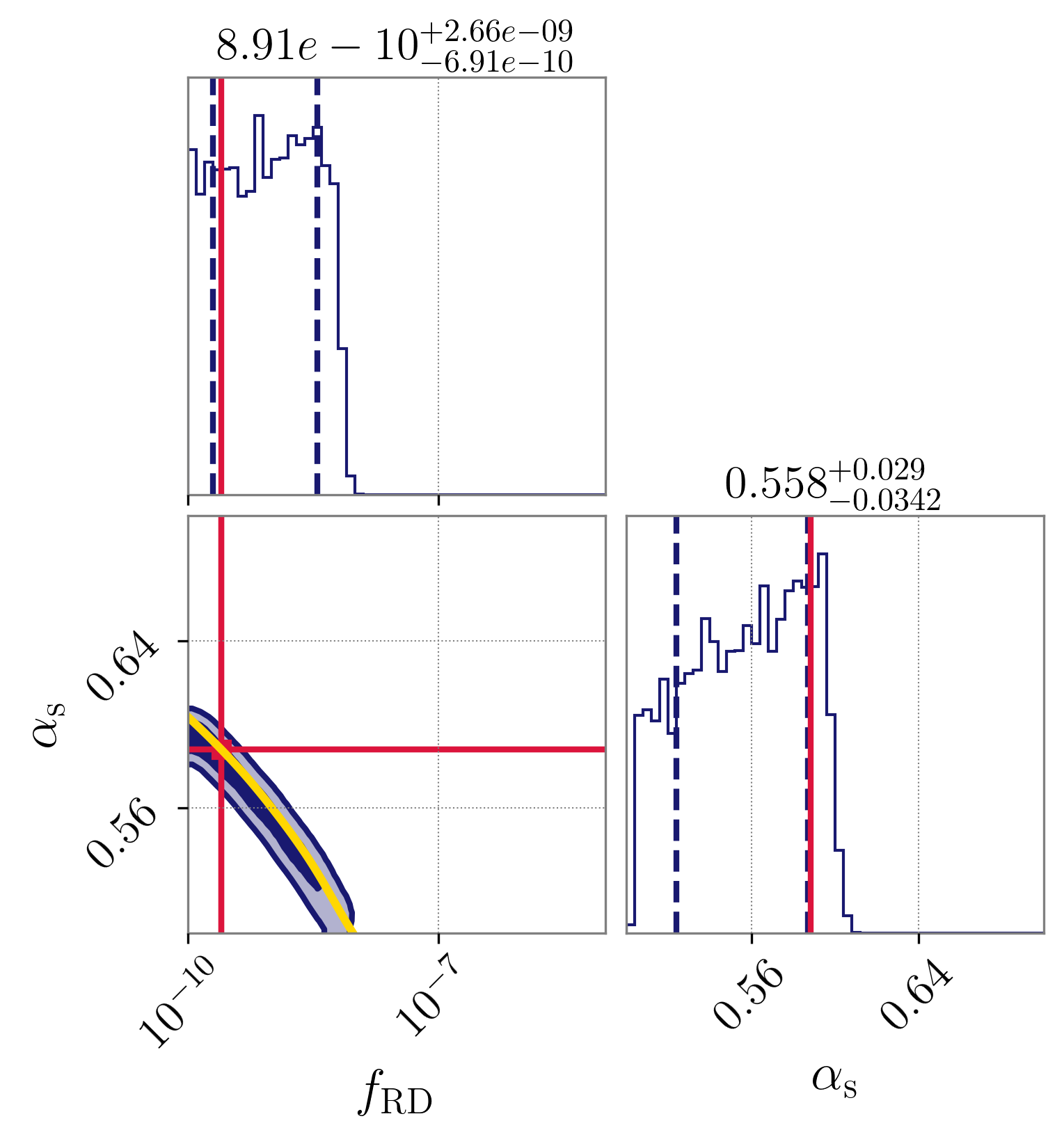}
  \caption{Comparison between Bayesian inference results and the analytically derived degeneracy in the $(f_{\rm RD}, \alpha_{\rm s})$ parameter space. The analytically derived degeneracy curve is shown in yellow.
  }
  \label{fig:PE_degeneracy_each}
  \hspace{-1cm}
\end{figure}

In conclusion, our analysis shows that a partial reconstruction of a possible signal from a stiff epoch will be feasible at Advanced LIGO A+ sensitivity, even though the presence of the CBC background and of the degeneracies in the predicted GWB  pose challenges to
a robust recovery of the relevant parameters.

\subsection{Detection prospects for the stiff epoch for future gravitational-wave experiments}
\label{sec:futuredetectors}
Here, we discuss the future detectability prospects across a wide parameter space, which is possible with future experiments that span various frequency bands
(see \cite{Co:2021lkc,Gouttenoire:2021wzu,gouttenoire2022kination} for previous similar studies for the case of kination, with further constraints arising in the axion implementation).
We assess the detectability for different values of $f_{\rm RD}$ and $f_{\rm SD}$ using the Advanced LIGO A+ sensitivity, LISA, ET, and a network of 3G detectors,
with corresponding power-law integrated (PLI) sensitivity curves\footnote{Although the signal follows a broken power law, we use the approximation of a power law to assess its detectability. Since the GW spectrum remains relatively smooth across the experimental frequency band, even at its peak, this approximation is well justified and commonly used in the literature, see e.g. \cite{Schmitz:2020syl, Badger:2022nwo}.  } 
given in Table~\ref{table:PI_curves}, for which we have assumed design sensitivities~\cite{ETB_Hild}. 
The sensitivity estimate used for ET is ET-D, the most updated version assuming a triangular configuration with arms of 10 km, obtained by incorporating additional noise sources ~\cite{Hild_2011}. The 3G network consists of four detectors: two CE-like detectors with arms of $40$ km, located in the current LIGO detectors' locations, and two ET-D-like detectors co-located and co-aligned at Virgo's location\footnote{For simplicity, in this study we have assumed ET and CE to be located at the current LIGO and Virgo locations.
However, their real locations are still under discussion. Furthermore, we work with the ET-D configuration, but there are different configurations possible for ET, see e.g. \cite{Branchesi:2023mws} for a recent study.}
}. Finally, for LISA we assume data processing with Time Delay Interferometry (TDI) variables and consider auto-correlation of two orthogonal signal channels, so called A- and E-channels that correspond to $+$-and $\times$-independent polarisation modes. 

We fix $w_{\rm s}$ to three different values: $w_{\rm s}=1.0$, $w_{\rm s}=0.8$ and $w_{\rm s}=0.6$,  corresponding to the maximum possible value of $w_{\rm s}$ (kination), and two middle ones. Considering smaller values of $w_{\rm s}$ leads to weaker GWB signals, and hence a narrower detectable parameter space.
The RD1-to-MD1 transition frequency $f_{\rm MD}$ and the inflationary scale $H_{\rm inf}$ are always fixed to their maximum values.
\begin{figure}[h!]
    \centering
    \includegraphics[scale=0.3]{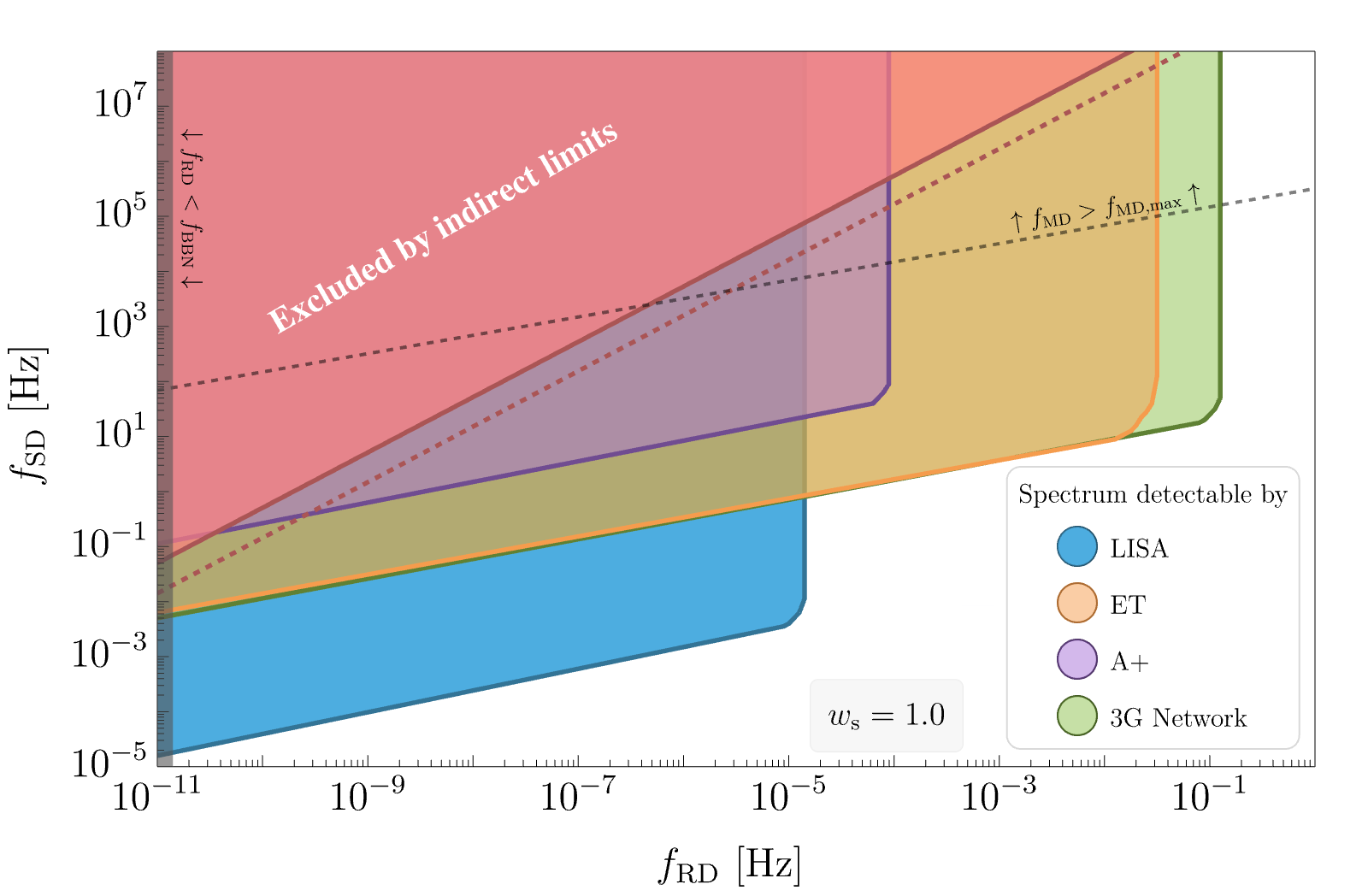}
    \includegraphics[scale=0.3]{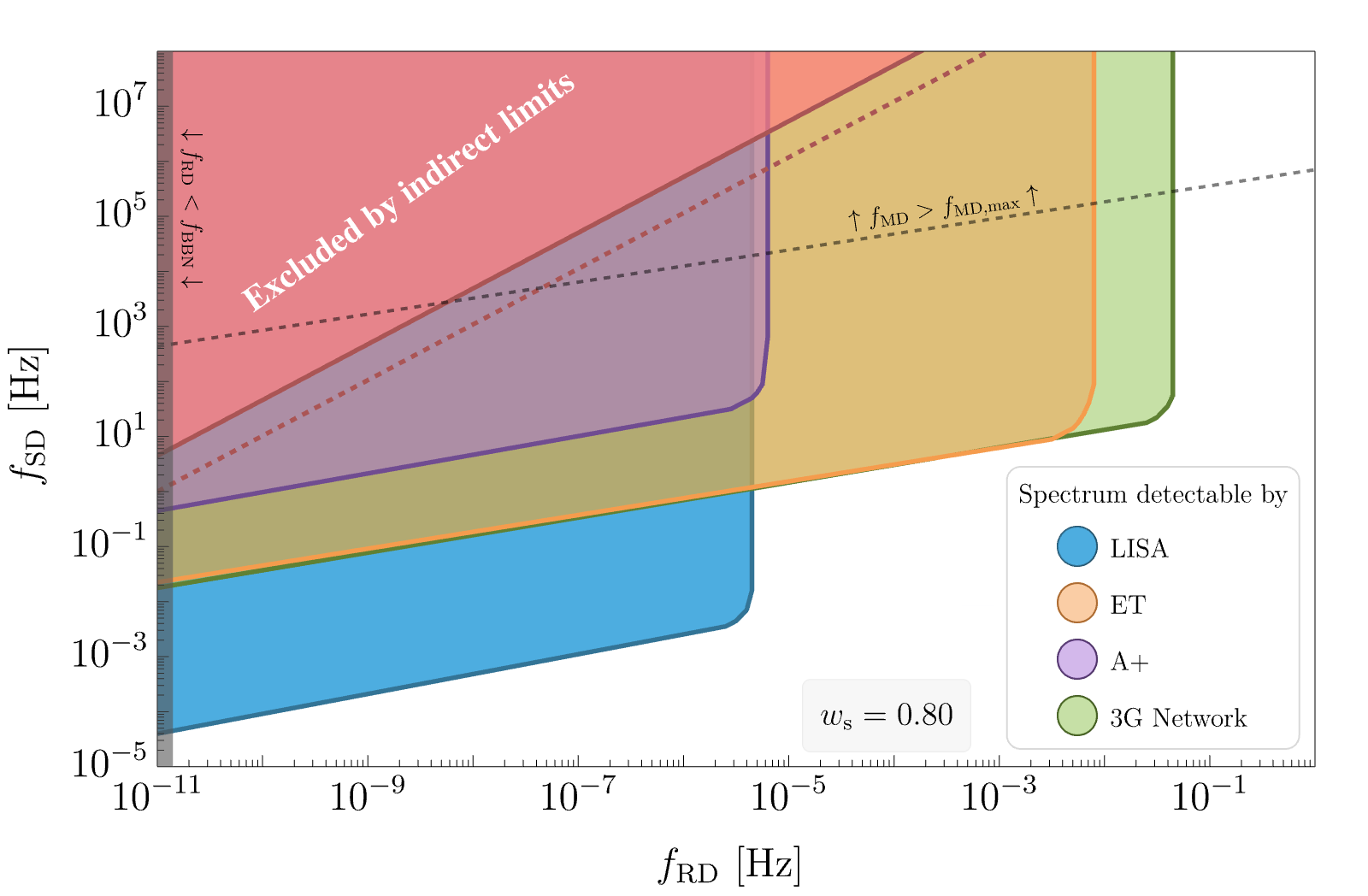}
    \includegraphics[scale=0.3]{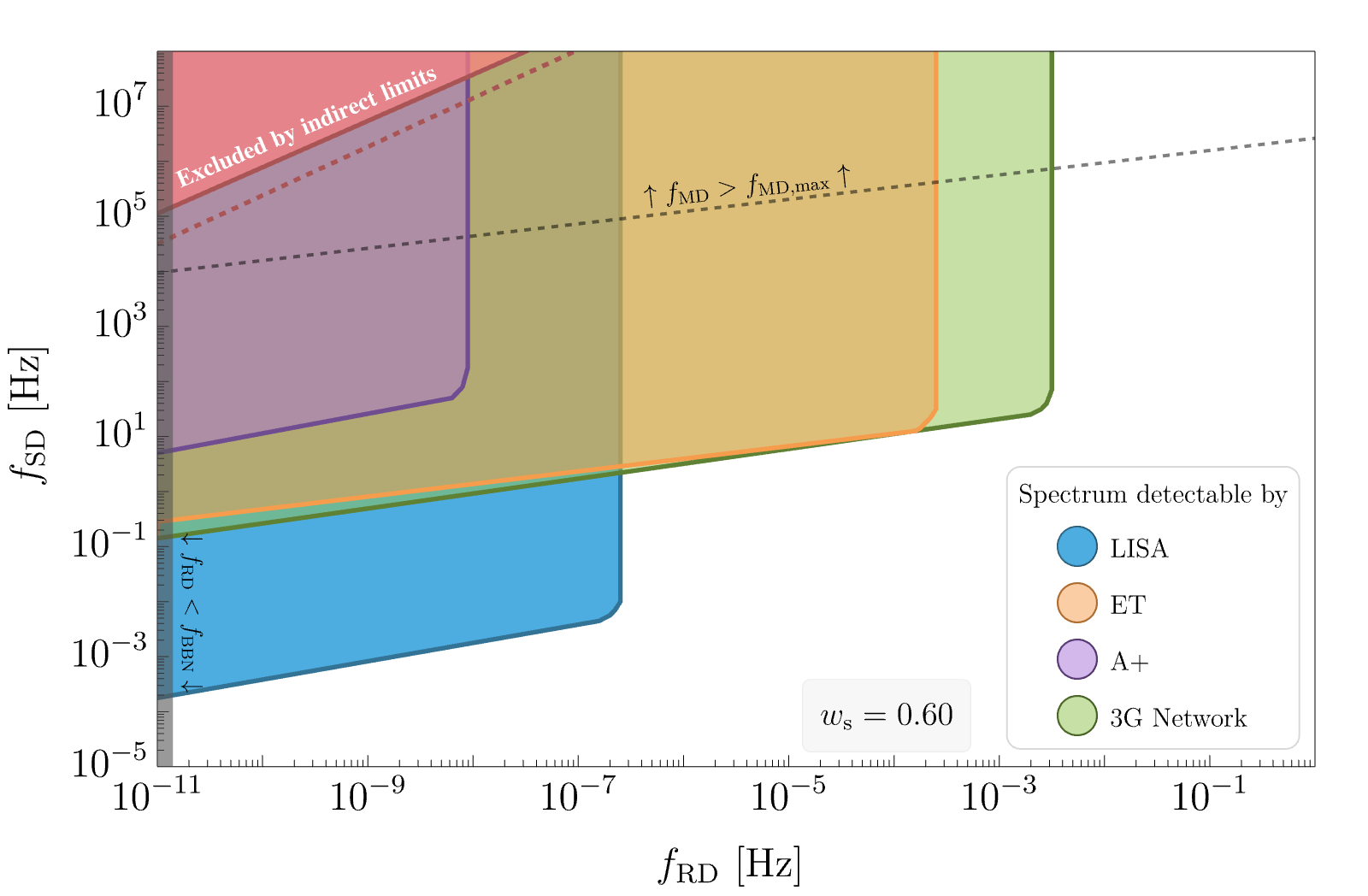}
    \caption{
    The ($f_{\rm RD}, f_{\rm SD}$) parameter regions accessible by future detectors are shown for three different values of $w_{\rm s}$ (from top to bottom: $w_{\rm s}=1.0$, $0.8$, and $0.6$). The coloured regions represent areas detectable by various future detectors: blue for LISA, orange for ET, purple for Advanced LIGO A+, and green for a network of 3G detectors.   
    The dashed line indicates $f_{\rm MD}=f_{\rm i}$, and the region above this line may require a modification of the cosmological history such that inflation ends in a MD era (see discussion after Eq. \eqref{eq:maxfmd}).
    The pink region gives the parameter space excluded by indirect limits from BBN and CMB, as described in Eq.~\eqref{eq:DeltaNeff}. The pink dashed line gives the future indirect limits region, described in Eq.~\eqref{eq:DeltaNefffuture}. 
    The grey area on the left 
    is excluded by the BBN constraint ($f_{\rm RD}\geq f_{\rm BBN}$).
    }
    \label{fig:paramspacebounds}
\end{figure}
As one can see from Fig.~\ref{fig:paramspacebounds}, a significant part of the ($f_{\rm RD}, f_{\rm SD}$) detectable space is excluded by indirect bounds, depicted in pink. 
This occurs when $f_{\rm SD}$ is large and hence the GW energy density exceeds the current $\Delta N_{\rm eff}$ bound (see Eq.~\eqref{eq:DeltaNeff}). 
However, there remains a substantial portion of allowed parameter space that is detectable by future experiments.
For larger values of $f_{\rm RD}$, the network of 3G detectors has the strongest reach, while for low values of $f_{\rm SD}$ LISA is the most promising. We also add future expected indirect bounds from $\Delta N_{\rm eff}$ as derived in Eq.~\eqref{eq:DeltaNefffuture}. 
The area above the black dashed line, denoted as $f_{\rm MD} > f_{\rm MD, max}$, represents the parameter region where $f_{\rm MD}$, as defined by Eq. \eqref{eq:frdMax}. 
exceeds its maximum value, defined in Eq. \eqref{eq:maxfmd}. 
We conclude that future GW experiments can significantly probe regions of the parameter space 
that would otherwise be
 untestable via indirect probes.

\begin{figure}[h!]
    \centering
    \includegraphics[scale=0.3]{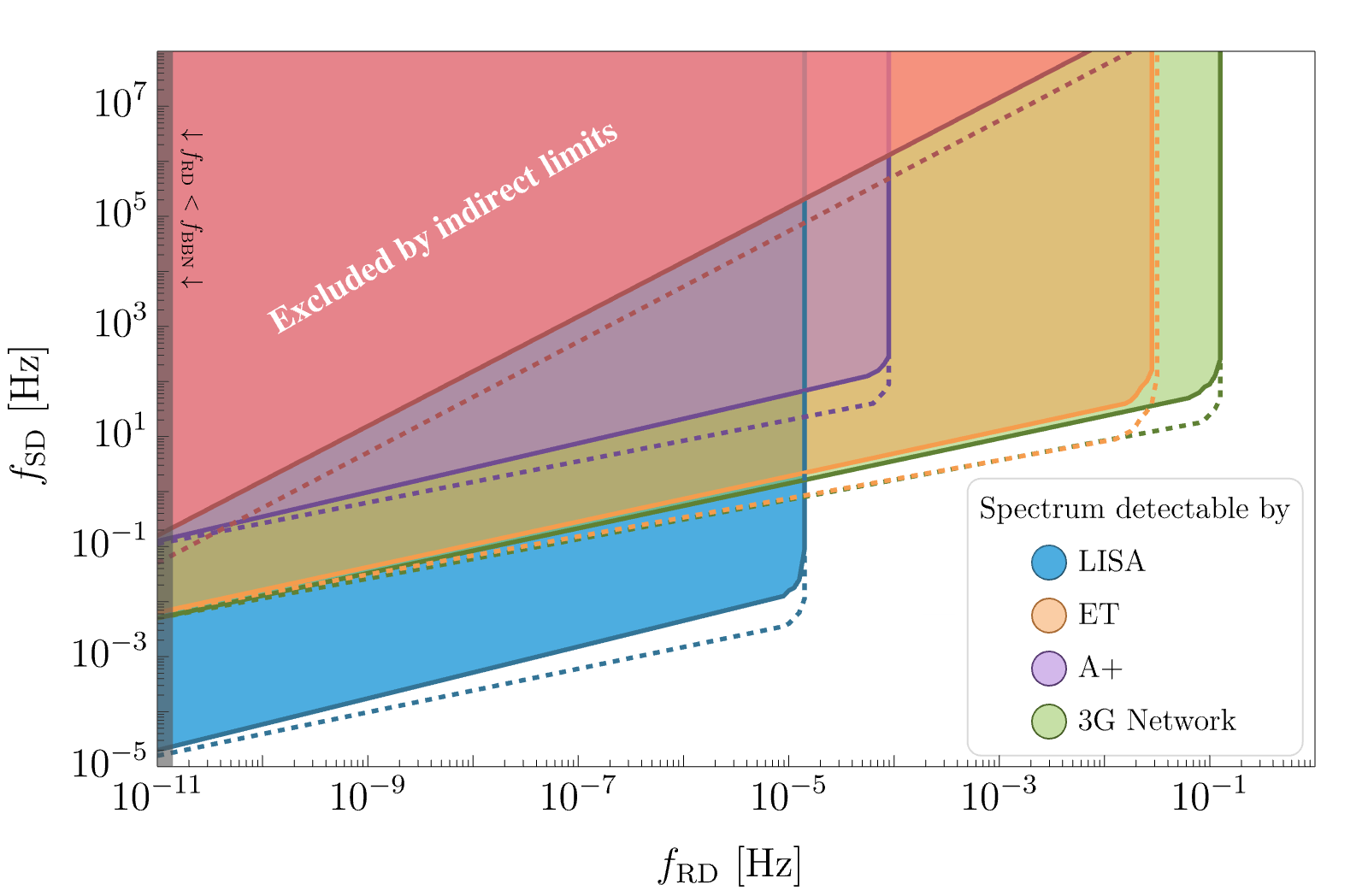}
    \caption{($f_{\rm RD}, f_{\rm SD}$) parameter regions for the smooth MD1-to-kination transition model, described in Section \ref{sec:smoothinstant}. The coloured regions correspond to regions that are detectable by different future detectors, where the different colours used are: blue for LISA, orange for ET, purple for A+ and green for a network of 3G detectors, and the dashed lines correspond to the instant MD1-to-kination model. The pink region gives the parameter space for the smooth case excluded by indirect limits from BBN and CMB, as described in Eq.~\eqref{eq:DeltaNeff}. The pink dashed line  gives the parameter space for the instant case excluded by indirect limits. The grey area on the left 
    is excluded by the BBN constraint ($f_{\rm RD}\geq f_{\rm BBN}$).
    }
    \label{fig:smoothinstantparamspaceplotzoom}
\end{figure}

We end this section by inspecting the differences in the detection prospects between 
the instant and the smooth case.
As was shown earlier, the GW peak of a smooth MD1-to-kination transition is different than the one of an instant transition. We investigate whether this difference leads to sizeable modifications in the  detectability prospects. In Fig.~\ref{fig:smoothinstantparamspaceplotzoom} we show the 
detectability regions in the ($f_{\rm RD}$, $f_{\rm SD}$) parameter space for the smooth GW spectrum given in Eq.~\eqref{eq:smoothGWspectrum}.
For comparison, we show the corresponding sensitivity regions for an instant MD-kination transition with dashed lines (which is the same case as in the top panel of Fig. ~\ref{fig:paramspacebounds}). We find that there only is a minor difference in the reach of  GW experiments. 
The difference arises because for 
a fixed $f_{\rm SD}$, the GW spectrum of the instant transition has a peak which is more pronounced and more shifted to high frequencies with respect to the smooth transition case.
However, it is important to note that the majority of the interesting regions for the future GW experiments is not modified.
Thus, for realistic models with smooth transitions, there would not be major modifications with respect to the findings assuming instant transitions.

\begin{table*}[htbp]
\centering 
\begin{tabular}{|| c | c | c | c | c | c ||}
 \hline
  PLI curve name & Description & Observation time &  SNR & Co-located and co-aligned & ASD curve \\ 
 \hline\hline
   A + & HL network, design sensitivities & 1 year & 1 & No &~\cite{Aplusdesign}\\ 
 \hline
  ET & 2 ET-D-like detectors & 1 year & 1 & Yes &~\cite{Hild_2011}\\
 \hline
  LISA & Auto-correlation of A- and E-channels & 1 year & 1 & - &~\cite{Caprini:2019pxz}\\
 \hline
  3G network & 2 ET-D-like detector and 2 CE-like detectors & 1 year & 1 & No &~\cite{Hild_2011},~\cite{CE-asd}\\
 \hline
\end{tabular}
\caption{Description of the PLI sensitivity curves used throughout this paper. 
The second column describes the configuration of each network. The third and fourth columns indicate the assumed observation time and the signal-to-noise ratio (SNR) used for the construction of these PLI curves. The fifth column specifies whether the detectors involved are co-located and co-aligned. Finally, the sixth column provides a reference for
the amplitude spectral density (ASD) noise curves used in constructing these PLI sensitivity curves. 
The A+ PLI curve consists of the Hanford-Livingston (HL) network.
The 3G network PLI curve consists of two ET-D-like-detectors co-located and co-aligned at the Virgo site, along with two CE-like detectors, each with 40 km arms, located at the current LIGO-Hanford and LIGO-Livingston sites. 
}
\label{table:PI_curves}
\end{table*}

\newpage
\section{Conclusion and discussion}
\label{sec:discussion}
We have considered a cosmological history, where an extra radiation dominated, a matter dominated and an era dominated by a stiff equation of state are present between the end of inflation and the onset of BBN. 
We have studied the influence of this unconventional cosmological history on the GWB sourced by inflation. 
The transitions between the different eras are described either as instantaneous (sudden) or smooth (gradual). 
The instant transitions serve as proxies for realistic models, enabling an analytical derivation of the GWB spectrum and capturing its phenomenology. Conversely, while the smooth transition scenario is more physically motivated, it demands numerical analysis.
We explore both types of transitions and work out the GWB in each case, resulting in a broken power law spectrum, with a distinctive peak.

We then perform a Bayesian inference analysis using O1-O3 LVK data. Even though our analysis shows no evidence for a GWB signal in the data, we can set 95\% CL ULs on some model parameters as well as on the amplitude of the CBC background, which is
consistent with the previously found upper limit ~\cite{Abbott_2021}.
In the case of kination ($w=1$),
the data can exclude scenarios where the value of the MD1-to-kination transition frequency is larger than $\sim 10$ Hz \emph{and} the kination-to-radiation transition is smaller than $\sim 10^{-5}$Hz (see the bottom panel of Fig. \ref{fig:PE_O3_1}). We also applied a Bayesian inference analysis using mock data assuming
Advanced LIGO A+ sensitivity 
and injecting
the GWB signals originating from a stiff epoch and of a CBC background.
Within this framework, various parameter degeneracies present 
challenges for extracting information on model parameters.

We have then investigated the impact of future GW experiments, specifically Advanced LIGO A+, LISA, ET, and a 3G network made of two ET-D-like detectors and two CE-like detectors. 
We have shown that there is a large parameter region that can be tested by future GW experiments and that is not in the reach of current and future indirect bounds. Finally, we have discussed the potential of the future GW experiments to differentiate between signals corresponding to instant or smooth transitions between the matter-to-stiff epochs. \\

There are several possible extensions to the work presented here.
It has been shown that second-order effects in GW perturbations can arise, depending on the speed of transitions between epochs, leading to additional features and enhancements in the GWB spectrum
(see e.g.~\cite{Inomata:2019ivs,Inomata:2020lmk,Domenech:2019quo,Harigaya:2023mhl,Pearce:2023kxp}).
We leave the investigation of these (model-dependent) effects and how they affect the detectability of the exotic cosmological history for the future.
In addition, we have neglected the backreaction of the GW energy density in the evolution of the Universe which, 
although typically small, could potentially result in additional indirect constraints
~\cite{Giovannini:1998bp,Li:2021htg}.

Throughout the paper,
we have assumed a model of inflation with vanishing primordial tilt ($n_{\rm t}=0$) in the tensor power spectrum.
We note that the GWB from the exotic cosmological history we considered, with $n_t=0$,
cannot accommodate the recent Pulsar Timing Array (PTA) observations
\cite{NANOGrav:2023gor,EPTA:2023fyk,Reardon:2023gzh,Xu:2023wog}
due to the lower limit on $f_{\rm RD} \geq f_{\rm BBN}$
(see e.g. ~\cite{Li:2021htg}).
This limitation prevents the spectrum from having a sufficient frequency range to increase at frequencies lower than the nano-Hertz band.
Instead, in scenarios with the described exotic cosmology \emph{and} a non-vanishing tilt, the inflationary GWB spectrum can reach the PTA sensitivity region\footnote{
Alternatively, one can explain the PTA data with a scalar-induced GW spectrum enhanced by an SD period~\cite{Harigaya:2023pmw}.
}~\cite{Kuroyanagi:2020sfw}.
The assumption of $n_t=0$ could be easily relaxed and our data-analysis could be repeated at the cost of adding an extra parameter (the spectral tilt), which would have a high degree of degeneracy with the equation of state during the stiff epoch. We leave this possibility for future investigations.

\section*{Acknowledgements}
We thank Daniel Figueroa, Sokratis Trifinopoulos, Kevin Turbang, Miguel Vanvlasselaer, and Santiago Jaraba for discussions. We are also grateful to Giancarlo Cella, Yue Zhao, Peera Simakachorn, Yann Gouttenoire, Géraldine Servant and Bruce Allen for useful comments on the draft. 

HD, AR and AM acknowledge support by the Strategic Research Program High-Energy Physics of the Research Council of the Vrije Universiteit Brussel, the iBOF ``Unlocking the Dark Universe with Gravitational Wave Observations: from Quantum Optics to Quantum Gravity" of the Vlaamse Interuniversitaire Raad, and the ``Excellence of Science - EOS" - be.h project n.30820817. SK is supported by the Spanish Atracci\'on de Talento contract no. 2019-T1/TIC-13177 granted by Comunidad de Madrid, the Spanish Research Agency (Agencia Estatal de Investigaci\'on) through the Grant IFT Centro de Excelencia Severo Ochoa No CEX2020-001007-S and the I+D grant PID2020-118159GA-C42 funded by MCIN/AEI/10.13039/501100011033, the Consolidaci\'on Investigadora 2022 grant CNS2022-135211 of CSIC, and Japan Society for the Promotion of Science (JSPS) KAKENHI Grant no. JP20H05853, and JP23H00110, JP24K00624. MS acknowledges support from the Science and Technology Facility Council
(STFC), UK, under the research grant ST/X000753/1. 

This research has made use of data or software obtained from the Gravitational Wave Open Science Center (\url{gwosc.org}), a service of the LIGO Scientific Collaboration, the Virgo Collaboration, and KAGRA. This material is based upon work supported by NSF's LIGO Laboratory which is a major facility fully funded by the National Science Foundation, as well as the Science and Technology Facilities Council (STFC) of the United Kingdom, the Max-Planck-Society (MPS), and the State of Niedersachsen/Germany for support of the construction of Advanced LIGO and construction and operation of the GEO600 detector. Additional support for Advanced LIGO was provided by the Australian Research Council. Virgo is funded, through the European Gravitational Observatory (EGO), by the French Centre National de Recherche Scientifique (CNRS), the Italian Istituto Nazionale di Fisica Nucleare (INFN) and the Dutch Nikhef, with contributions by institutions from Belgium, Germany, Greece, Hungary, Ireland, Japan, Monaco, Poland, Portugal, Spain. KAGRA is supported by the Ministry of Education, Culture, Sports, Science and Technology (MEXT), Japan Society for the Promotion of Science (JSPS) in Japan; National Research Foundation (NRF) and Ministry of Science and ICT (MSIT) in Korea; Academia Sinica (AS) and National Science and Technology Council (NSTC) in Taiwan.

The software packages we used are {\tt matplotlib}~\cite{4160265}, {\tt numpy}~\cite{van_der_Walt_2011}, {\tt scipy}~\cite{2020NatMe..17..261V}, {\tt bilby}~\cite{Ashton_2019} and {\tt pygwb}~\cite{Renzini_2023}.

This article has a LIGO document number LIGO-P2400169 and a Virgo document number VIR-0379B-24.

\appendix

\section{Analytical derivation of the gravitational-wave spectrum}\label{appendixmatching}
We show the formalism needed to analytically derive the GW spectrum, when assuming instant transitions of the equation of state. The solutions to the GW EOM during \textit{era 1}, which is followed by \textit{era 2}, are given by Eq.~\eqref{eq: GWsolint}. 
A non-trivial task that rests us is to determine the coefficients $A$ and $B$ for each epoch. 
This is achieved by matching the solutions and their time derivatives to those 
obtained for
the subsequent epoch at the transition moment $\tau_{1 \rightarrow 2}$ (see~\cite{Figueroa_2019} for the inflation-SD-RD transition case). These matching conditions are expressed as
\begin{align}
\begin{cases}
    h^{\rm epoch1 } (\tau_{\rm 1 \rightarrow 2}) &= h^{\rm epoch 2} (\tau_{\rm  1 \rightarrow 2}) \\
    h^{'{\rm epoch 1}} (\tau_{\rm 1 \rightarrow 2}) &= h^{'\rm epoch 2} (\tau_{\rm 1 \rightarrow 2})~,
\end{cases}
\label{eq:matching}
\end{align}
where 
\begin{align*}
    h^{'{\rm epoch 1}} (\tau_{\rm 1 \rightarrow 2}) = \frac{1}{\alpha_1} \frac{dh^{\rm epoch 1}(y)}{d y}\Big|_{y = \kappa_{\rm  1\rightarrow 2}}\,,
\end{align*}
with $\kappa_{\rm  1 \rightarrow 2}\equiv k/k_{\rm  1 \rightarrow 2} \equiv k/(a_{\rm  1 \rightarrow 2} H_{\rm  1 \rightarrow 2})$. Additionally, the specific value of $\alpha_1$ depends on the nature of \textit{era 1}, for instance, for RD eras $\alpha_1=1$, for MD eras $\alpha_1=2$ and for SD eras $\alpha_1=\alpha_{\rm s}$.\\

The only matching that does not follow Eq. \eqref{eq:matching}, is the first matching we perform, at the transition moment $\tau_{\rm i}$ between inflation and the RD1 era. For this, we impose the following boundary conditions
\begin{equation}
\begin{cases}
    h^{\rm RD1}(\kappa_{{\rm i}}) &= h_{\rm inf} \text{ \ for \ } \kappa_{{\rm i}} \ll 1 \\
    \frac{d h^{\rm RD1}(y)}{dy}\Big|_{y=\kappa_{\rm i}} &= 0 \text{ \ for\ } \kappa_{{\rm i}} \ll 1 
\end{cases}
\end{equation}
where $\kappa_{\rm i} \equiv k/(a_{\rm i} H_{\rm i})$ and
\begin{equation}
    h_{\rm inf} \equiv \frac{1}{\sqrt{k^3}}\left(\frac{H_{\rm inf}}{m_{\rm Pl}}\right)~.
\end{equation}
Therefore, we are able to find the solution of Eq.~\eqref{eq:GWeqn} during the RD1 era:
\begin{align}
    h^{\rm RD1} (y) = 2^{\frac{1}{2}} h_{\rm inf} \Gamma\left(\frac{3}{2}\right) y^{-\frac{1}{2}} J_{\frac{1}{2}} (y)  ~.
\end{align}
For the next epochs, we impose the continuity of the tensor modes and their derivatives, as given in Eq. ~\eqref{eq:matching}.
We apply these matching conditions for the RD1-to-MD1 transition at $\tau_{\rm MD}$, for the MD1-to-SD transition at $\tau_{\rm SD}$ and for the SD-to-RD2 transition at $\tau_{\rm RD}$.
In this Appendix, we work out one example: the matching at RD1-to-MD1 transition. In MD1, the solution is given by
\begin{align}
    h^{\rm MD 1}(y) &= A_{\rm MD1} (2 y)^{-\frac{3}{2}} J_{\frac{3}{2}}(2 y) \\
    &+ B_{\rm MD1} (2 y)^{-\frac{3}{2}} Y_{\frac{3}{2}}(2 y) \,,
\end{align}
We then demand that 
\begin{align*}
    \begin{cases}
    h^{\rm RD1} (\kappa_{\rm MD}) &= h^{\rm MD} (\kappa_{\rm MD}) \\
    \frac{d h^{\rm RD1} (y)}{d y}\Big|_{y=\kappa_{\rm MD}} &= \frac{1}{2} \frac{d h^{\rm RD1} (y)}{d y}\Big|_{y=\kappa_{\rm MD}}
\end{cases}
\end{align*}
The coefficients are then found to be equal to, in terms of $\kappa_{MD} = k/k_{MD}= y_{MD}$:
\small
\begin{align*}
\begin{cases}
    A_{\rm MD1} ( \kappa_{\rm MD}) \\
    B_{\rm MD1} ( \kappa_{\rm MD})
\end{cases}
= 2 h_{\rm inf} \Gamma \left(\frac{3}{2}\right) \times \frac{1}{c_{\rm MD1}(\kappa_{\rm MD})}
\begin{cases}
    a_{\rm MD1} ( \kappa_{\rm MD}) \\
    b_{\rm MD1} ( \kappa_{\rm MD}) 
\end{cases}
~,
\end{align*}
\normalsize
with
\begin{align*}
    c_{\rm MD1}( \kappa_{\rm MD}) =&\left( J_{\frac{5}{2}}(2 \kappa_{\rm MD})-J_{\frac{1}{2}}(2 \kappa_{\rm MD})\right) Y_{\frac{3}{2}}(2 \kappa_{\rm MD}) \\
   &+J_{\frac{3}{2}}(2 \kappa_{\rm MD}) \left(Y_{\frac{1}{2}}(2 \kappa_{\rm MD})-Y_{\frac{5}{2}}(2
   \kappa_{\rm MD})\right) ~,
\end{align*} 
\small
\begin{align*}
    &a_{\rm MD1}( \kappa_{\rm MD}) = \left(2
   \kappa_{\rm MD} \left(J_{-\frac{1}{2}}(\kappa_{\rm MD})-J_{\frac{3}{2}}(\kappa_{\rm MD})\right)
   Y_{\frac{3}{2}}(2 \kappa_{\rm MD}) \right.\\
   &\left. +J_{\frac{1}{2}}(\kappa_{\rm MD}) \left(-2 \kappa_{\rm MD}
   Y_{\frac{1}{2}}(2 \kappa_{\rm MD})+Y_{\frac{3}{2}}(2 \kappa_{\rm MD})+2 \kappa_{\rm MD} Y_{\frac{5}{2}}(2 \kappa_{\rm MD})\right)\right) 
\end{align*}
and 
\begin{align*}
    &b_{\rm MD1}( \kappa_{\rm MD}) =\left(2
   \kappa_{\rm MD} \left(J_{ -\frac{1}{2}}(\kappa_{\rm MD})-J_{\frac{3}{2}}(\kappa_{\rm MD})\right) J_{\frac{3}{2}}(2 \kappa_{\rm MD}) \right.\\
   &\left. +J_{\frac{1}{2}}(\kappa_{\rm MD}) \left(-2 \kappa_{\rm MD}
   J_{\frac{1}{2}}(2 \kappa_{\rm MD})+J_{\frac{3}{2}}(2 \kappa_{\rm MD})+2 \kappa_{\rm MD} J_{\frac{5}{2}}(2 \kappa_{\rm MD})\right)\right) ~.
\end{align*}
\normalsize
Repeating this procedure for the MD1-to-SD transition and for the SD-to-RD2 transition, we are able to construct the GW spectrum
\begin{equation}
\label{eq:OmegaGW}
    \Omega_{\rm GW}(\tau,k) = \frac{k^2}{12 a^2(\tau) H^2(\tau)}\Delta_{\rm h}^2(\tau,k)~,
\end{equation}
where the present-day tensor power spectrum is
\begin{equation}
    \Delta_{\rm h}^2(\tau_0,k) = \frac{2k^3}{\pi^2} \overline{(h^{\rm RD 2} (\tau_0)^2)}~.
\end{equation}
To construct the present GW spectrum, we need to work with the RD2 solution $h^{\rm RD 2}(y)$, which contains all information on the previous eras encoded in the coefficients $A_{\rm RD 2}(\kappa_{\rm MD}, \kappa_{\rm SD},  \kappa_{\rm RD}, \alpha_{\rm s}) $ and $B_{\rm RD 2}(\kappa_{\rm MD}, \kappa_{\rm SD},  \kappa_{\rm RD}, \alpha_{\rm s}) $ and is given by 
\begin{equation}
    h^{\rm RD 2} (y) = \sqrt{\frac{2}{\pi}} \frac{A_{\rm RD 2}\sin(y) - B_{\rm RD 2}\cos(y)}{y}~.
\end{equation}
By squaring and averaging over its mode oscillations, we get
\begin{equation}
    \overline{\big(h^{\rm RD2} (y)\big)^2} = \frac{A_{\rm RD2}^2 + B_{\rm RD2}^2}{\pi} y^2~,
\end{equation} 
Using $(\frac{H_{\rm RD}}{H_0})^2 (\frac{a_{\rm RD}}{a_0})^4 = \Omega_{\rm rad}^{(0)} \mathcal{G}_k$, the final GW spectrum can be found to be 
\begin{eqnarray}
    \Omega_{\rm GW}(\tau_0,k) =&& \frac{4^{-2} G_k H_{\rm inf}^2 \Omega_{\rm rad}^{(0)} \Gamma \left(\frac{3}{2}\right)^2}{3 \pi ^3 \alpha_{\rm s} ^4 m_{\rm Pl}^2}\nonumber \\
    &&\times
   \left(\frac{f}{f_{\rm RD}}\right)^{-2 \alpha} \left(\frac{f}{f_{\rm SD}}\right)^{2 (\alpha_{\rm s} -3)} \nonumber\\
   &&\times  {\cal F}(f, f_{\rm RD}, f_{\rm SD}, f_{\rm MD}, \alpha_{\rm s})~,
\end{eqnarray}
where ${\cal F}(f, f_{\rm RD}, f_{\rm SD}, f_{\rm MD}, \alpha_{\rm s})$ is a combination of Bessel functions. For conciseness, we refrain from explicitly stating its expression here. However, it can be derived through the procedure outlined above\footnote{We give the full expression of the GWB here: \url{https://github.com/HanDuval/GWB-from-stiff-epoch/tree/6ef78504b13bef2491385ff580e0101bf4b0bad9}.}.

\section{Logarithmic potential for a smooth transition}
\label{app:LogPot}
\begin{figure}[h!]
    \includegraphics[width=8cm]{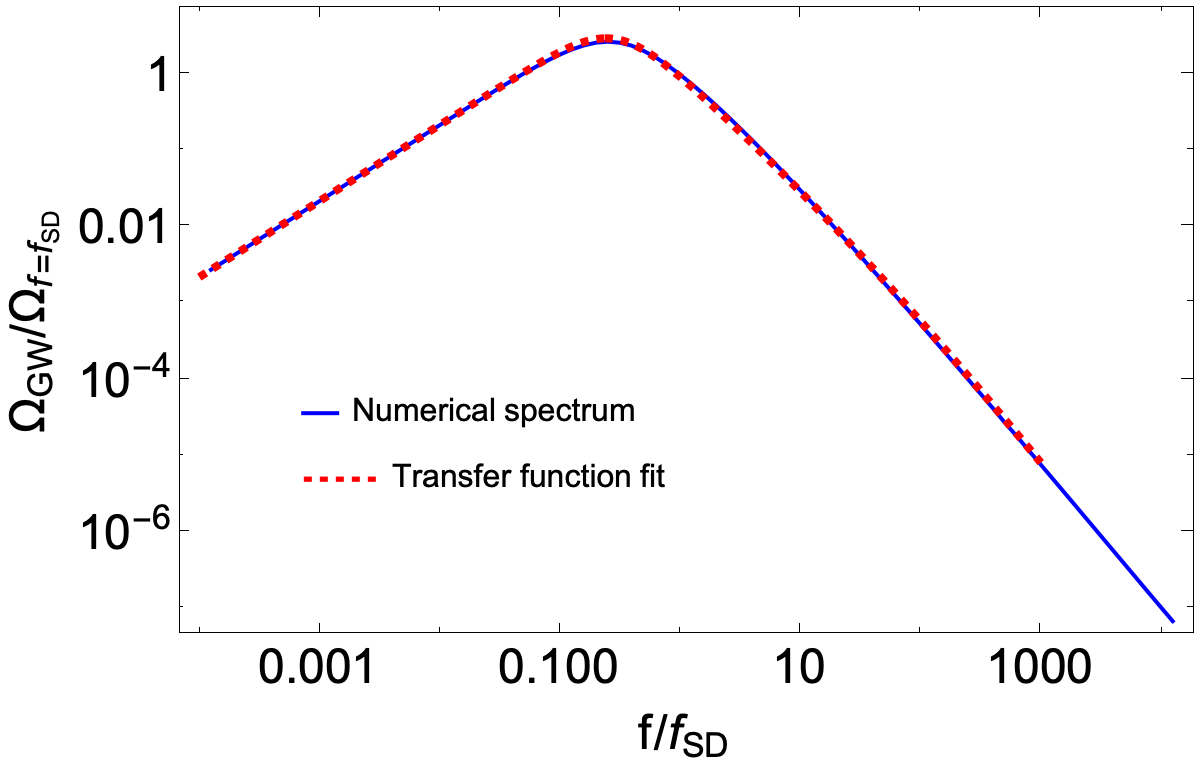}
    \caption{The 
    numerically derived GWB spectrum, shown in blue, and the GWB fitted using the transfer function method, depicted in red. The  horizontal and vertical axes are normalised by $f_{\rm SD}$ and $\Omega_{\rm GW}(f_{\rm SD})$ respectively.}
    \label{fig:smoothGW}
\end{figure}

We review the results of 
\cite{Co:2021lkc} and show how we obtained the transfer function associated to the smooth case scenario.
We consider the model proposed in~\cite{Co:2019wyp,Co:2021lkc,gouttenoire2022kination, Gouttenoire:2021wzu}
where a simple realisation of a period of kination is studied by considering a complex scalar field with a logarithmic potential
\begin{equation}
V = m_{\rm S}^2 |P|^2 \left[\log \left(\frac{2 |P|^2}{f_a^2}  \right) -1 \right] + \frac{1}{2}m_{\rm S}^2 f_a^2~.
\end{equation}
The complex field is split in a radial mode $S$ and an angle $\theta$, and $P$ is defined as $P=(S/\sqrt{2}) e^{i \theta/f_a}$.
The EOM for $\theta$ implies the conservation of the angular momentum,
$\dot \theta S^2 \sim a^{-3}$ in an expanding Universe.
To start, we assume that at large field value for the radial mode $S \gg f_a$, the evolution goes from a matter equation of state to kination.
The precise shape of the transition can be obtained by solving the EOM in the approximation $\dot S \sim \ddot S \sim 0$ 
leading to
\begin{equation}
\label{Sanal}
S^2 = 
\frac{2 a_i^3 S_i^2}{a^3} \sqrt{\frac{\log \left(\frac{S_i}{f_a}\right)}{W\left[\frac{4 a_i^6 S_i^4 \log \left(\frac{S_i}{f_a}\right)}{a^6 f_a^4}\right]}}~,
\end{equation}
where $S_i$ is the initial field value, with $S_i \gg f_a$, $a_i$ is the initial scale factor,
and $W$ is the Lambert function.
The logarithmic derivative of the energy density is:
\\
\\
\begin{equation}
\frac{d \log \rho}{d \log a}=
-\frac{6 S^2 \log \left(\frac{S^2}{f_a^2}\right)}{f_a^2 \left(-\frac{S^2}{f_a^2}+\frac{2 S^2 \log \left(\frac{S^2}{f_a^2}\right)}{f_a^2}+1\right)}~.
\end{equation}
The evolution of the energy density goes from
MD (when $S\gg f_a$) to kination (when $S=f_a$). 
One then solves the GW EOM numerically for a Universe dominated by the energy density of this complex field, and obtains the spectrum as shown in Fig.~\ref{fig:smoothGW}.
In this figure, we also show the fitted transfer function we have derived and reported in the main text of the paper. 
It accurately reproduces the numerical GW spectrum.
Note that the frequency $f_{\rm SD}$ is defined by requiring that asymptotically, for large $f$, the spectrum goes as
$\left(f/f_{\rm SD}\right)^{-3} (\mathcal{A}_2/\mathcal{A}_{1/2})
$.
Since the smooth transition reaches the MD era only asymptotically, 
this procedure cannot unambiguously fix $f_{\rm SD}$.
We verified that our choice does not lead to modifications of more than a few percentages on the resulting GWB peak.

Note that, given that the energy density evolution in the smooth transition is not symmetric on the two sides of the transition, the resulting peak is not aligned with the frequency $f_{\rm SD}$.

\newpage
\bibliography{bibliography} 

\begin{thebibliography}{10}

\bibitem{PhysRevX.11.021053}
{The LIGO Scientific Collaboration}, {The Virgo Collaboration}, {The KAGRA Collaboration}, R.~Abbott, {\em et~al.}, ``{GWTC-2}: Compact binary coalescences observed by {LIGO} and {Virgo} during the first half of the third observing run,'' {\em Phys. Rev. X}, vol.~11, p.~021053, Jun 2021.

\bibitem{PhysRevX.13.041039}
{The LIGO Scientific Collaboration}, {The Virgo Collaboration}, {The KAGRA Collaboration}, R.~Abbott, {\em et~al.}, ``{GWTC-3}: Compact binary coalescences observed by {LIGO} and {Virgo} during the second part of the third observing run,'' {\em Phys. Rev. X}, vol.~13, p.~041039, Dec 2023.

\bibitem{Allen_Romano_99}
B.~Allen and J.~D. Romano, ``Detecting a stochastic background of gravitational radiation: Signal processing strategies and sensitivities,'' {\em Phys. Rev. D}, vol.~59, p.~102001, Mar 1999.

\bibitem{Romano_Cornish_2017}
J.~Romano and N.~Cornish, ``Detection methods for stochastic gravitational-wave backgrounds: A unified treatment,'' {\em Living Reviews in Relativity}, vol.~20, 04 2017.

\bibitem{Caprini:2018mtu}
C.~Caprini and D.~G. Figueroa, ``{Cosmological Backgrounds of Gravitational Waves},'' {\em Class. Quant. Grav.}, vol.~35, no.~16, p.~163001, 2018.

\bibitem{Regimbau:2011rp}
T.~Regimbau, ``{The astrophysical gravitational wave stochastic background},'' {\em Res. Astron. Astrophys.}, vol.~11, pp.~369--390, 2011.

\bibitem{Starobinsky:1979ty}
A.~A. Starobinsky, ``{Spectrum of relict gravitational radiation and the early state of the universe},'' {\em JETP Lett.}, vol.~30, pp.~682--685, 1979.

\bibitem{Rubakov:1982df}
V.~A. Rubakov, M.~V. Sazhin, and A.~V. Veryaskin, ``{Graviton Creation in the Inflationary Universe and the Grand Unification Scale},'' {\em Phys. Lett. B}, vol.~115, pp.~189--192, 1982.

\bibitem{Giovannini:1998bp}
M.~Giovannini, ``{Gravitational waves constraints on postinflationary phases stiffer than radiation},'' {\em Phys. Rev. D}, vol.~58, p.~083504, 1998.

\bibitem{Peebles:1998qn}
P.~J.~E. Peebles and A.~Vilenkin, ``{Quintessential inflation},'' {\em Phys. Rev. D}, vol.~59, p.~063505, 1999.

\bibitem{Giovannini:1999bh}
M.~Giovannini, ``{Production and detection of relic gravitons in quintessential inflationary models},'' {\em Phys. Rev. D}, vol.~60, p.~123511, 1999.

\bibitem{Giovannini:1999qj}
M.~Giovannini, ``{Spikes in the relic graviton background from quintessential inflation},'' {\em Class. Quant. Grav.}, vol.~16, pp.~2905--2913, 1999.

\bibitem{Giovannini:2008tm}
M.~Giovannini, ``{Thermal history of the plasma and high-frequency gravitons},'' {\em Class. Quant. Grav.}, vol.~26, p.~045004, 2009.

\bibitem{Tashiro:2003qp}
H.~Tashiro, T.~Chiba, and M.~Sasaki, ``{Reheating after quintessential inflation and gravitational waves},'' {\em Class. Quant. Grav.}, vol.~21, pp.~1761--1772, 2004.

\bibitem{Figueroa:2018twl}
D.~G. Figueroa and E.~H. Tanin, ``{Inconsistency of an inflationary sector coupled only to Einstein gravity},'' {\em JCAP}, vol.~10, p.~050, 2019.

\bibitem{Figueroa_2019}
D.~G. Figueroa and E.~H. Tanin, ``Ability of {LIGO} and {LISA} to probe the equation of state of the early universe,'' {\em Journal of Cosmology and Astroparticle Physics}, vol.~2019, pp.~011--011, aug 2019.

\bibitem{Co:2019wyp}
R.~T. Co and K.~Harigaya, ``{Axiogenesis},'' {\em Phys. Rev. Lett.}, vol.~124, no.~11, p.~111602, 2020.

\bibitem{Co:2021lkc}
R.~T. Co, D.~Dunsky, N.~Fernandez, A.~Ghalsasi, L.~J. Hall, K.~Harigaya, and J.~Shelton, ``{Gravitational wave and {CMB} probes of axion kination},'' {\em JHEP}, vol.~09, p.~116, 2022.

\bibitem{gouttenoire2022kination}
Y.~Gouttenoire, G.~Servant, and P.~Simakachorn, ``Kination cosmology from scalar fields and gravitational-wave signatures,'' 2022.

\bibitem{Gouttenoire:2021wzu}
Y.~Gouttenoire, G.~Servant, and P.~Simakachorn, ``{Revealing the Primordial Irreducible Inflationary Gravitational-Wave Background with a Spinning Peccei-Quinn Axion},'' 8 2021.

\bibitem{Seto:2003kc}
N.~Seto and J.~Yokoyama, ``{Probing the equation of state of the early universe with a space laser interferometer},'' {\em J. Phys. Soc. Jap.}, vol.~72, pp.~3082--3086, 2003.

\bibitem{Smith:2005mm}
T.~L. Smith, M.~Kamionkowski, and A.~Cooray, ``{Direct detection of the inflationary gravitational wave background},'' {\em Phys. Rev. D}, vol.~73, p.~023504, 2006.

\bibitem{Boyle_2008}
L.~A. Boyle and A.~Buonanno, ``Relating gravitational wave constraints from primordial nucleosynthesis, pulsar timing, laser interferometers, and the cmb: Implications for the early universe,'' {\em Phys. Rev. D}, vol.~78, p.~043531, Aug 2008.

\bibitem{Nakayama:2008ip}
K.~Nakayama, S.~Saito, Y.~Suwa, and J.~Yokoyama, ``{Space laser interferometers can determine the thermal history of the early Universe},'' {\em Phys. Rev. D}, vol.~77, p.~124001, 2008.

\bibitem{Nakayama:2008wy}
K.~Nakayama, S.~Saito, Y.~Suwa, and J.~Yokoyama, ``{Probing reheating temperature of the universe with gravitational wave background},'' {\em JCAP}, vol.~06, p.~020, 2008.

\bibitem{Nakayama:2009ce}
K.~Nakayama and J.~Yokoyama, ``{Gravitational Wave Background and Non-Gaussianity as a Probe of the Curvaton Scenario},'' {\em JCAP}, vol.~01, p.~010, 2010.

\bibitem{Kuroyanagi:2008ye}
S.~Kuroyanagi, T.~Chiba, and N.~Sugiyama, ``{Precision calculations of the gravitational wave background spectrum from inflation},'' {\em Phys. Rev. D}, vol.~79, p.~103501, 2009.

\bibitem{Kuroyanagi:2011fy}
S.~Kuroyanagi, K.~Nakayama, and S.~Saito, ``{Prospects for determination of thermal history after inflation with future gravitational wave detectors},'' {\em Phys. Rev. D}, vol.~84, p.~123513, 2011.

\bibitem{Kuroyanagi:2013ns}
S.~Kuroyanagi, C.~Ringeval, and T.~Takahashi, ``{Early universe tomography with {CMB} and gravitational waves},'' {\em Phys. Rev. D}, vol.~87, no.~8, p.~083502, 2013.

\bibitem{Kuroyanagi:2014qza}
S.~Kuroyanagi, K.~Nakayama, and J.~Yokoyama, ``{Prospects of determination of reheating temperature after inflation by DECIGO},'' {\em PTEP}, vol.~2015, no.~1, p.~013E02, 2015.

\bibitem{Kuroyanagi:2014nba}
S.~Kuroyanagi, T.~Takahashi, and S.~Yokoyama, ``{Blue-tilted Tensor Spectrum and Thermal History of the Universe},'' {\em JCAP}, vol.~02, p.~003, 2015.

\bibitem{Bernal:2019lpc}
N.~Bernal and F.~Hajkarim, ``{Primordial Gravitational Waves in Nonstandard Cosmologies},'' {\em Phys. Rev. D}, vol.~100, no.~6, p.~063502, 2019.

\bibitem{Bernal:2020ywq}
N.~Bernal, A.~Ghoshal, F.~Hajkarim, and G.~Lambiase, ``{Primordial Gravitational Wave Signals in Modified Cosmologies},'' {\em JCAP}, vol.~11, p.~051, 2020.

\bibitem{DEramo:2019tit}
F.~D'Eramo and K.~Schmitz, ``{Imprint of a scalar era on the primordial spectrum of gravitational waves},'' {\em Phys. Rev. Research.}, vol.~1, p.~013010, 2019.

\bibitem{Li:2013nal}
B.~Li, T.~Rindler-Daller, and P.~R. Shapiro, ``{Cosmological Constraints on Bose-Einstein-Condensed Scalar Field Dark Matter},'' {\em Phys. Rev. D}, vol.~89, no.~8, p.~083536, 2014.

\bibitem{Li:2016mmc}
B.~Li, P.~R. Shapiro, and T.~Rindler-Daller, ``{Bose-Einstein-condensed scalar field dark matter and the gravitational wave background from inflation: new cosmological constraints and its detectability by {LIGO}},'' {\em Phys. Rev. D}, vol.~96, no.~6, p.~063505, 2017.

\bibitem{Li:2021htg}
B.~Li and P.~R. Shapiro, ``{Precision cosmology and the stiff-amplified gravitational-wave background from inflation: NANOGrav, Advanced {LIGO-Virgo} and the {Hubble} tension},'' {\em JCAP}, vol.~10, p.~024, 2021.

\bibitem{Mishra:2021wkm}
S.~S. Mishra, V.~Sahni, and A.~A. Starobinsky, ``{Curing inflationary degeneracies using reheating predictions and relic gravitational waves},'' {\em JCAP}, vol.~05, p.~075, 2021.

\bibitem{Battefeld:2004jh}
T.~J. Battefeld and D.~A. Easson, ``{Perturbations in a holographic universe and in other stiff fluid cosmologies},'' {\em Phys. Rev. D}, vol.~70, p.~103516, 2004.

\bibitem{DeAngelis:2024xtr}
M.~De~Angelis, A.~Smith, W.~Giar\`e, and C.~van~de Bruck, ``{Gravitational waves in a cyclic Universe: resilience through cycles and vacuum state},'' 3 2024.

\bibitem{Haque:2021dha}
M.~R. Haque, D.~Maity, T.~Paul, and L.~Sriramkumar, ``{Decoding the phases of early and late time reheating through imprints on primordial gravitational waves},'' {\em Phys. Rev. D}, vol.~104, no.~6, p.~063513, 2021.

\bibitem{Chakraborty:2023ocr}
A.~Chakraborty, M.~R. Haque, D.~Maity, and R.~Mondal, ``{Inflaton phenomenology via reheating in light of primordial gravitational waves and the latest BICEP/Keck data},'' {\em Phys. Rev. D}, vol.~108, no.~2, p.~023515, 2023.

\bibitem{LIGOScientific:2016jlg}
B.~P. Abbott {\em et~al.}, ``{Upper Limits on the Stochastic Gravitational-Wave Background from Advanced {LIGO}\textquoteright{}s First Observing Run},'' {\em Phys. Rev. Lett.}, vol.~118, no.~12, p.~121101, 2017.
\newblock [Erratum: Phys.Rev.Lett. 119, 029901 (2017)].

\bibitem{LIGOScientific:2019vic}
B.~P. Abbott {\em et~al.}, ``{Search for the isotropic stochastic background using data from Advanced {LIGO}\textquoteright{}s second observing run},'' {\em Phys. Rev. D}, vol.~100, no.~6, p.~061101, 2019.

\bibitem{KAGRA:2021kbb}
R.~Abbott {\em et~al.}, ``{Upper limits on the isotropic gravitational-wave background from {Advanced LIGO} and Advanced {Virgo}\textquoteright{}s third observing run},'' {\em Phys. Rev. D}, vol.~104, no.~2, p.~022004, 2021.

\bibitem{Aplusdesign}
L.~Barsotti, L.~McCuller, M.~Evans, and P.~Fritschel, ``The {A+} design curve,'' {\em LIGO Document T1800042-v5}, Jan 2020.

\bibitem{colpi2024lisa}
M.~Colpi {\em et~al.}, ``{LISA} definition study report,'' 2024.

\bibitem{Punturo:2010zz}
M.~Punturo {\em et~al.}, ``{The {Einstein Telescope}: A third-generation gravitational wave observatory},'' {\em Class. Quant. Grav.}, vol.~27, p.~194002, 2010.

\bibitem{Hild_2011}
S.~Hild, M.~Abernathy, F.~Acernese, P.~Amaro-Seoane, N.~Andersson, K.~Arun, F.~Barone, B.~Barr, M.~Barsuglia, M.~Beker, and et. al., ``Sensitivity studies for third-generation gravitational wave observatories,'' {\em Classical and Quantum Gravity}, vol.~28, p.~094013, apr 2011.

\bibitem{borhanian2024listening}
S.~Borhanian and B.~S. Sathyaprakash, ``Listening to the universe with next generation ground-based gravitational-wave detectors,'' 2024.

\bibitem{Guzzetti:2016mkm}
M.~C. Guzzetti, N.~Bartolo, M.~Liguori, and S.~Matarrese, ``{Gravitational waves from inflation},'' {\em Riv. Nuovo Cim.}, vol.~39, no.~9, pp.~399--495, 2016.

\bibitem{Ade_2021}
P.~A.~R. Ade, Z.~Ahmed, R.~Aikin, and K.~e.~a. Alexander, ``Improved constraints on primordial gravitational waves using planck, wmap, and bicep/keck observations through the 2018 observing season,'' {\em Physical Review Letters}, vol.~127, Oct. 2021.

\bibitem{Turner:1993vb}
M.~S. Turner, M.~J. White, and J.~E. Lidsey, ``{Tensor perturbations in inflationary models as a probe of cosmology},'' {\em Phys. Rev. D}, vol.~48, pp.~4613--4622, 1993.

\bibitem{Turner:1996ck}
M.~S. Turner, ``{Detectability of inflation produced gravitational waves},'' {\em Phys. Rev. D}, vol.~55, pp.~R435--R439, 1997.

\bibitem{Turner:1983he}
M.~S. Turner, ``{Coherent Scalar Field Oscillations in an Expanding Universe},'' {\em Phys. Rev. D}, vol.~28, p.~1243, 1983.

\bibitem{Watanabe:2006qe}
Y.~Watanabe and E.~Komatsu, ``{Improved Calculation of the Primordial Gravitational Wave Spectrum in the Standard Model},'' {\em Phys. Rev. D}, vol.~73, p.~123515, 2006.

\bibitem{Saikawa:2018rcs}
K.~Saikawa and S.~Shirai, ``{Primordial gravitational waves, precisely: The role of thermodynamics in the Standard Model},'' {\em JCAP}, vol.~05, p.~035, 2018.

\bibitem{PhysRevD.37.2078}
B.~Allen, ``Stochastic gravity-wave background in inflationary-universe models,'' {\em Phys. Rev. D}, vol.~37, pp.~2078--2085, Apr 1988.

\bibitem{Litebird_2022}
E.~Allys, K.~Arnold, J.~Aumont, R.~Aurlien, and e.~a. Azzoni, ``Probing cosmic inflation with thelitebirdcosmic microwave background polarization survey,'' {\em Progress of Theoretical and Experimental Physics}, vol.~2023, Nov. 2022.

\bibitem{Caprini_2018}
C.~Caprini and D.~G. Figueroa, ``Cosmological backgrounds of gravitational waves,'' {\em Classical and Quantum Gravity}, vol.~35, p.~163001, jul 2018.

\bibitem{Yeh_2022}
T.-H. Yeh, J.~Shelton, K.~A. Olive, and B.~D. Fields, ``Probing physics beyond the standard model: limits from bbn and the {CMB} independently and combined,'' {\em Journal of Cosmology and Astroparticle Physics}, vol.~2022, p.~046, Oct. 2022.

\bibitem{abazajian2016cmbs4}
K.~N. Abazajian, P.~Adshead, Z.~Ahmed, S.~W. Allen, D.~Alonso, {\em et~al.}, ``{CMB-S4} science book, first edition,'' 2016.

\bibitem{SimonsObservatory:2018koc}
P.~Ade {\em et~al.}, ``{The Simons Observatory: Science goals and forecasts},'' {\em JCAP}, vol.~02, p.~056, 2019.

\bibitem{LIGOScientific:2019lzm}
R.~Abbott {\em et~al.}, ``{Open data from the first and second observing runs of Advanced {LIGO} and Advanced {Virgo}},'' {\em SoftwareX}, vol.~13, p.~100658, 2021.

\bibitem{KAGRA:2023pio}
R.~Abbott {\em et~al.}, ``{Open Data from the Third Observing Run of {LIGO}, {Virgo}, {KAGRA}, and {GEO}},'' {\em Astrophys. J. Suppl.}, vol.~267, no.~2, p.~29, 2023.

\bibitem{PhysRevLett.109.171102}
V.~Mandic, E.~Thrane, S.~Giampanis, and T.~Regimbau, ``Parameter estimation in searches for the stochastic gravitational-wave background,'' {\em Phys. Rev. Lett.}, vol.~109, p.~171102, Oct 2012.

\bibitem{PhysRevX.7.041058}
T.~Callister, A.~S. Biscoveanu, N.~Christensen, M.~Isi, A.~Matas, O.~Minazzoli, T.~Regimbau, M.~Sakellariadou, J.~Tasson, and E.~Thrane, ``Polarization-based tests of gravity with the stochastic gravitational-wave background,'' {\em Phys. Rev. X}, vol.~7, p.~041058, Dec 2017.

\bibitem{PhysRevD.102.102005}
P.~M. Meyers, K.~Martinovic, N.~Christensen, and M.~Sakellariadou, ``Detecting a stochastic gravitational-wave background in the presence of correlated magnetic noise,'' {\em Phys. Rev. D}, vol.~102, p.~102005, Nov 2020.

\bibitem{Abbott_2021}
R.~Abbott, T.~D. Abbott, and S.~e.~a. Abraham, ``Upper limits on the isotropic gravitational-wave background from advanced {LIGO} and advanced {Virgo’s} third observing run,'' {\em Physical Review D}, vol.~104, July 2021.

\bibitem{PhysRevD.59.102001}
B.~Allen and J.~D. Romano, ``Detecting a stochastic background of gravitational radiation: Signal processing strategies and sensitivities,'' {\em Phys. Rev. D}, vol.~59, p.~102001, Mar 1999.

\bibitem{Renzini_2023}
A.~I. Renzini, A.~Romero-Rodríguez, C.~Talbot, M.~Lalleman, S.~Kandhasamy, K.~Turbang, S.~Biscoveanu, K.~Martinovic, P.~Meyers, L.~Tsukada, K.~Janssens, D.~Davis, A.~Matas, P.~Charlton, G.-C. Liu, I.~Dvorkin, S.~Banagiri, S.~Bose, T.~Callister, F.~D. Lillo, L.~D’Onofrio, F.~Garufi, G.~Harry, J.~Lawrence, V.~Mandic, A.~Macquet, I.~Michaloliakos, S.~Mitra, K.~Pham, R.~Poggiani, T.~Regimbau, J.~D. Romano, N.~van Remortel, and H.~Zhong, ``pygwb: A python-based library for gravitational-wave background searches,'' {\em The Astrophysical Journal}, vol.~952, p.~25, jul 2023.

\bibitem{Ashton_2019}
G.~Ashton, M.~Hübner, P.~D. Lasky, C.~Talbot, K.~Ackley, S.~Biscoveanu, Q.~Chu, A.~Divakarla, P.~J. Easter, B.~Goncharov, F.~H. Vivanco, J.~Harms, M.~E. Lower, G.~D. Meadors, D.~Melchor, E.~Payne, M.~D. Pitkin, J.~Powell, N.~Sarin, R.~J.~E. Smith, and E.~Thrane, ``Bilby: A user-friendly bayesian inference library for gravitational-wave astronomy,'' {\em The Astrophysical Journal Supplement Series}, vol.~241, p.~27, Apr. 2019.

\bibitem{Abbott_2018}
B.~P. Abbott, R.~Abbott, and T.~D. e.~a. Abbott, ``Gw170817: Implications for the stochastic gravitational-wave background from compact binary coalescences,'' {\em Physical Review Letters}, vol.~120, Feb. 2018.

\bibitem{Martinovic_2021}
K.~Martinovic, P.~M. Meyers, M.~Sakellariadou, and N.~Christensen, ``Simultaneous estimation of astrophysical and cosmological stochastic gravitational-wave backgrounds with terrestrial detectors,'' {\em Physical Review D}, vol.~103, Feb. 2021.

\bibitem{Schmitz:2020syl}
K.~Schmitz, ``{New Sensitivity Curves for Gravitational-Wave Signals from Cosmological Phase Transitions},'' {\em JHEP}, vol.~01, p.~097, 2021.

\bibitem{Badger:2022nwo}
C.~Badger {\em et~al.}, ``{Probing early Universe supercooled phase transitions with gravitational wave data},'' {\em Phys. Rev. D}, vol.~107, no.~2, p.~023511, 2023.

\bibitem{ETB_Hild}
S.~Hild, S.~Chelkowski, and A.~Freise, ``Pushing towards the et sensitivity using 'conventional' technology,'' 11 2008.

\bibitem{Branchesi:2023mws}
M.~Branchesi {\em et~al.}, ``{Science with the Einstein Telescope: a comparison of different designs},'' {\em JCAP}, vol.~07, p.~068, 2023.

\bibitem{Caprini:2019pxz}
C.~Caprini, D.~G. Figueroa, R.~Flauger, G.~Nardini, M.~Peloso, M.~Pieroni, A.~Ricciardone, and G.~Tasinato, ``{Reconstructing the spectral shape of a stochastic gravitational wave background with {LISA}},'' {\em JCAP}, vol.~11, p.~017, 2019.

\bibitem{CE-asd}
K.~Kuns, E.~Hall, V.~Srivastava, J.~Smith, M.~Evans, P.~Fritschel, L.~McCuller, C.~Wipf, and S.~Ballmer, ``Cosmic explorer strain sensitivity,'' {\em CE Document CE-T2000017-v7}, Jan 2021.

\bibitem{Inomata:2019ivs}
K.~Inomata, K.~Kohri, T.~Nakama, and T.~Terada, ``{Enhancement of Gravitational Waves Induced by Scalar Perturbations due to a Sudden Transition from an Early Matter Era to the Radiation Era},'' {\em Phys. Rev. D}, vol.~100, p.~043532, 2019.
\newblock [Erratum: Phys.Rev.D 108, 049901 (2023)].

\bibitem{Inomata:2020lmk}
K.~Inomata, M.~Kawasaki, K.~Mukaida, T.~Terada, and T.~T. Yanagida, ``{Gravitational Wave Production right after a Primordial Black Hole Evaporation},'' {\em Phys. Rev. D}, vol.~101, no.~12, p.~123533, 2020.

\bibitem{Domenech:2019quo}
G.~Dom\`enech, ``{Induced gravitational waves in a general cosmological background},'' {\em Int. J. Mod. Phys. D}, vol.~29, no.~03, p.~2050028, 2020.

\bibitem{Harigaya:2023mhl}
K.~Harigaya, K.~Inomata, and T.~Terada, ``{Gravitational wave production from axion rotations right after a transition to kination},'' {\em Phys. Rev. D}, vol.~108, no.~8, p.~L081303, 2023.

\bibitem{Pearce:2023kxp}
M.~Pearce, L.~Pearce, G.~White, and C.~Bal\'azs, ``{Gravitational Wave Signals From Early Matter Domination: Interpolating Between Fast and Slow Transitions},'' 11 2023.

\bibitem{NANOGrav:2023gor}
G.~Agazie {\em et~al.}, ``{The NANOGrav 15 yr Data Set: Evidence for a Gravitational-wave Background},'' {\em Astrophys. J. Lett.}, vol.~951, no.~1, p.~L8, 2023.

\bibitem{EPTA:2023fyk}
J.~Antoniadis {\em et~al.}, ``{The second data release from the European Pulsar Timing Array - III. Search for gravitational wave signals},'' {\em Astron. Astrophys.}, vol.~678, p.~A50, 2023.

\bibitem{Reardon:2023gzh}
D.~J. Reardon {\em et~al.}, ``{Search for an Isotropic Gravitational-wave Background with the Parkes Pulsar Timing Array},'' {\em Astrophys. J. Lett.}, vol.~951, no.~1, p.~L6, 2023.

\bibitem{Xu:2023wog}
H.~Xu {\em et~al.}, ``{Searching for the Nano-Hertz Stochastic Gravitational Wave Background with the Chinese Pulsar Timing Array Data Release I},'' {\em Res. Astron. Astrophys.}, vol.~23, no.~7, p.~075024, 2023.

\bibitem{Harigaya:2023pmw}
K.~Harigaya, K.~Inomata, and T.~Terada, ``{Induced gravitational waves with kination era for recent pulsar timing array signals},'' {\em Phys. Rev. D}, vol.~108, no.~12, p.~123538, 2023.

\bibitem{Kuroyanagi:2020sfw}
S.~Kuroyanagi, T.~Takahashi, and S.~Yokoyama, ``{Blue-tilted inflationary tensor spectrum and reheating in the light of NANOGrav results},'' {\em JCAP}, vol.~01, p.~071, 2021.

\bibitem{4160265}
J.~D. Hunter, ``Matplotlib: A 2d graphics environment,'' {\em Computing in Science \& Engineering}, vol.~9, no.~3, pp.~90--95, 2007.

\bibitem{van_der_Walt_2011}
S.~van~der Walt, S.~C. Colbert, and G.~Varoquaux, ``The numpy array: A structure for efficient numerical computation,'' {\em Computing in Science \& Engineering}, vol.~13, p.~22–30, Mar. 2011.

\bibitem{2020NatMe..17..261V}
P.~{Virtanen}, R.~{Gommers}, T.~E. {Oliphant}, M.~{Haberland}, T.~{Reddy}, D.~{Cournapeau}, E.~{Burovski}, P.~{Peterson}, W.~{Weckesser}, J.~{Bright}, S.~J. {van der Walt}, M.~{Brett}, J.~{Wilson}, K.~J. {Millman}, N.~{Mayorov}, A.~R.~J. {Nelson}, E.~{Jones}, R.~{Kern}, E.~{Larson}, C.~J. {Carey}, {\.I}.~{Polat}, Y.~{Feng}, E.~W. {Moore}, J.~{VanderPlas}, D.~{Laxalde}, J.~{Perktold}, R.~{Cimrman}, I.~{Henriksen}, E.~A. {Quintero}, C.~R. {Harris}, A.~M. {Archibald}, A.~H. {Ribeiro}, F.~{Pedregosa}, P.~{van Mulbregt}, and {SciPy 1. 0 Contributors}, ``{SciPy 1.0: fundamental algorithms for scientific computing in Python},'' {\em Nature Methods}, vol.~17, pp.~261--272, Feb. 2020.

\end{thebibliography}

\end{document}